\documentclass[graybox]{svmult}

\usepackage{hyperref}
\usepackage[utf8]{inputenc}
\usepackage{multirow}
\usepackage{mathptmx}       
\usepackage{helvet}         
\usepackage{courier}        
\usepackage{type1cm}        
\usepackage{makeidx}         
\usepackage{graphicx}        
\usepackage{multicol}        
\usepackage[bottom]{footmisc}
\usepackage[T1]{fontenc}
\makeindex             
\usepackage{amsmath}
\usepackage{mhchem}
\usepackage[mathscr]{eucal}
\usepackage{bm}
\usepackage{pifont}
\makeatletter
\renewcommand*\env@matrix[1][\arraystretch]{%
  \edef\arraystretch{#1}%
  \hskip -\arraycolsep
  \let\@ifnextchar\new@ifnextchar
  \array{*\c@MaxMatrixCols c}}
\makeatother
\DeclareMathOperator{\Tr}{Tr}

\graphicspath{ {./pictures/}, {./pictures/orbitals/}}

\usepackage{physics}
\usepackage{nicefrac}
\usepackage[numbers]{natbib}

\begin{document}

\title*{New Strategies in Modeling Electronic Structures and Properties with Applications to Actinides}
\titlerunning{New Strategies in Modeling Electronic Structures} 
\author{Aleksandra \L{}achma\'{n}ska, Pawe\l{} Tecmer, and Katharina Boguslawski}
\institute{Aleksandra \L{}achma\'{n}ska
\at Institute of Physics, Faculty of Physics, Astronomy and Informatics, Nicolaus Copernicus University in Torun, Grudziadzka 5, 87-100 Torun, Poland, 
\and Pawe\l{} Tecmer 
\at Institute of Physics, Faculty of Physics, Astronomy and Informatics, Nicolaus Copernicus University in Torun, Grudziadzka 5, 87-100 Torun, Poland, 
\and Katharina Boguslawski
\at Faculty of Chemistry, Nicolaus Copernicus University in Torun, Gagarina 7, 87-100 Torun, Poland, and   
\at Institute of Physics, Faculty of Physics, Astronomy and Informatics, Nicolaus Copernicus University in Torun, Grudziadzka 5, 87-100 Torun, Poland, 
\email{k.boguslawski@fizyka.umk.pl}
}
%
%
\maketitle

\abstract*{In this chapter we briefly review contemporary computational methods that find application in heavy-element chemistry.}

\abstract{This chapter discusses contemporary quantum chemical methods and provides general insights into modern electronic structure theory with a focus on heavy-element-containing compounds. 
We first give a short overview of relativistic Hamiltonians that are frequently applied to account for relativistic effects.
Then, we scrutinize various quantum chemistry methods that approximate the $N$-electron wave function.
In this respect, we will review the most popular single- and multi-reference approaches that have been developed to model the multi-reference nature of heavy element compounds and their ground- and excited-state electronic structures.
Specifically, we introduce various flavors of post-Hartree--Fock methods and optimization schemes like the complete active space self-consistent field method, the configuration interaction approach, the Fock-space coupled cluster model, the pair-coupled cluster doubles ansatz, also known as the antisymmetric product of 1 reference orbital geminal, and the density matrix renormalization group algorithm. 
Furthermore, we will illustrate how concepts of quantum information theory provide us with a qualitative understanding of complex electronic structures using the picture of interacting orbitals.
While modern quantum chemistry facilitates a quantitative description of atoms and molecules as well as their properties, concepts of quantum information theory offer new strategies for a qualitative interpretation that can shed new light onto the chemistry of complex molecular compounds.}
\section{Introduction}
\label{sec:1}
One of the main goals of quantum chemistry is to understand the physicochemical properties of atoms, molecules, and materials using the first principles. 
This knowledge can be further used to interpret and explain existing experimental data or to design new compounds with much sought-after properties.  
However, the molecules under investigation usually contain numerous interacting electrons, which leads to a complex computational problem with a large number of degrees of freedom. 
The interplay between relativistic effects, the correlated motion of electrons, and the basis set quality is the main difficulty that limits the possibility to express the electronic wave function in exact form.
Various quantum chemical methods have been successfully applied to molecular systems where these effects play a minor role.
However, molecules containing heavy elements like actinides or other d- and f-block elements still pose a challenge to quantum chemistry as both correlation and relativistic effects have a dominant contribution to their electronic structure.

In this chapter, we review conventional and unconventional quantum chemical theories that are applicable to heavy-element chemistry like actinide-containing compounds.
Our discussion starts with presenting the properties of actinides as an example of complex many-electron systems.
Then, we briefly summarize some popular approaches that account for relativistic effects, followed by electronic structure methods that optimize (approximate) electronic wave functions for ground and excited states.
Furthermore, we outline how information from the electronic wave function can be extracted to obtain a qualitative interpretation of electronic structures.
Specifically, our analysis covers concepts of quantum information theory.
Finally, we present some challenging examples of computational actinide chemistry that highlight the difficulty in describing the electronic structure of actinide-containing compounds.

\section{A Brief Overview of Actinides and Their Complex Electronic Structure}
Heavy elements with atomic numbers ranging from 89 to 103 form a distinct group in the periodic table known as actinides. 
This series includes actinium, the early actinides (thorium, protactinium, uranium, and neptunium), the middle actinides (plutonium, americium, curium, berkelium, and californium), and the late actinides (einsteinium, fermium, mendelevium, nobelium, and lawrencium). 
All elements are radioactive metals and almost all of them are characterized by short lifetimes. Only some isotopes of thorium and uranium elements have long lifetimes and thus can be found in nature.
Thorium, uranium, neptunium, plutonium, americium, and curium have important applications in the nuclear industry, whereas thorium and uranium are also exploited in catalysis.
%

In actinide elements, the 5$f$ electrons strongly interact with the remaining valence electrons as well as with each other. 
This interaction implicates an electronic structure composed of many quasi-degenerate electronic configurations. 
Examples are the 5$f^n$7$s^2$ or 5$f^{n-1}$6$d$7$s^2$ series of electronic configurations, where $n$ represents the number of electrons in the 5$f$ shell and is defined as $Z-88$, with $Z$ being the atomic number. 
These numerous energetically close-lying electronic configurations are, however, of different character across the actinide series, which causes irregularities in the electronic ground and excited state energies of actinide elements. 

Similar to transition metals, the highest (formal) oxidation state of the early actinides equals the total number of electrons that can be removed from the valence shell, that is, from the 6$d$ and 5$f$ atomic orbitals. 
Furthermore, the early actinides resemble transition metals also in terms of orbitals and valence properties. 
The main reason for the close resemblance of actinides and transition metals is that the actinide 6$d$ orbitals do participate in chemical bonding with other elements~\cite{Straka-actinide-t-metals-2005,Mikheev2007}. 
Recently, Wilson \textit{et al.}~\cite{Valerie-nature-2018} observed the energetic crossing of the 5$f$ and 6$d$ atomic states for protactinium, making the protactinium atom a potential crossing point of valence properties that are characteristic for either transition metals or actinides.  
Using quantum chemistry methods, the authors provided numerical evidence that both the 5$f$ and 6$d$ orbitals participate in the chemistry of Pa and that the participation of the 5$f$ orbitals increases for the middle actinides. 

Unfortunately, experimental manipulations with actinide species are very limited, primarily because most actinide atoms are unstable, feature a large number of various oxidation states, or are radiotoxic. 
Despite these technical difficulties, experimental actinide chemistry remains an active field of research that mainly focuses on molecular synthesis of compounds containing thorium and uranium as well as spectroscopic studies of such compounds~\cite{exp_goncharov_ie,denning2007,Arnold2009,fortier10,Heaven_UO+_06}.
Due to these difficulties, theoretical modeling of actinide-con\-taining compounds can complement experimental studies and provide the much sought-after insights into the physico-chemical properties of actinide complexes and clusters, their reaction mechanisms, and thermochemistry.
However, theoretical modeling of actinide chemistry is challenging for present-day quantum chemistry as our theoretical model has to account for (i) relativistic effects and (ii) the correlated motion of electrons.

Due to the large atomic number present in actinide atoms, relativistic effects considerably affect the electronic structure of actinide-containing compounds and may change the character of the principle configuration compared to calculations where relativistic effects are ignored. 
For instance, the relativistic mass correction to the core electrons causes the contraction of their corresponding orbital radii, while the valence orbitals are expanded leading to elongated chemical bonds.~\cite{autschbach2002}
Furthermore, spin-orbit interactions, which are comparable in magnitude to the electron-electron repulsion energy, reduce the degeneracies of states with non-zero angular momentum~\cite{ola-neptunyl-cci-2018}.
To appropriately model the correlated motion of the electrons, our electronic structure method has to include all degenerate or quasi-degenerate, low-lying electronic configurations resulting from the energetic proximity of the actinide valence 5$f$, 6$d$, and 7$s$ orbitals.
Such calculations are usually rather expensive.
Hence, various approximations have been introduced in quantum chemistry that allow us to efficiently treat (quasi-)degeneracies.
\section{Electronic Structure Methods in Quantum Chemistry}
\label{sec:2}

In the standard formulation of quantum chemistry, the quantum state of atoms and molecules consisting of $N$ electrons and $M$ nuclei is described by the total wave function $\Psi(\boldsymbol{x}, \boldsymbol{R})$, which depends on the spatial and spin coordinates $\boldsymbol{x} \equiv \{ \boldsymbol{x}_1, \boldsymbol{x}_2, ..., \boldsymbol{x}_N$\} of all electrons as well as on the spatial coordinates of all nuclei $\boldsymbol{R} \equiv \{ \boldsymbol{R}_1, \boldsymbol{R}_2, ..., \boldsymbol{R}_M\}$.
In quantum chemistry, we are usually interested in the electronic part of the wave function at a given molecular geometry, for instance the equilibrium structure.
Within the so-called Born-Oppenheimer approximation, the total wave function is then written as a product of a nuclear part and an electronic part.
In particular, the electronic part of the total wave function $\Psi_{\rm el}(\boldsymbol{x}; \boldsymbol{R})$ depends on all electronic coordinates, while the positions of the nuclei remain fixed and enter the wave function as parameters.
In non-relativistic quantum chemistry, the electronic wave function is obtained by solving the time-independent, electronic Schr\"{o}dinger equation
\begin{equation}
{H}_{\rm el} \Psi_{\rm el}(\boldsymbol{x}; \boldsymbol{R}) = E_{\rm el} \Psi_{\rm el}(\boldsymbol{x}; \boldsymbol{R}),
\end{equation}
where $H_{\rm el}$ denotes the Hamiltonian of the system, whose eigenvalues $E_{\rm el}$ are the electronic energies.
Typically, the non-relativistic electronic Hamiltonian $H_{\rm el}$ of a molecular system containing $N$ electrons and $M$ nuclei is given in Hartree atomic units ($\hbar = m_e = 4\pi\varepsilon_0 = 1$) and reads
\begin{equation}\label{eq:schroedingerH}
{H}_{\rm el} = - \sum_{i=1}^{N} \frac{1}{2} \nabla_{i}^{2}  - \sum_{i=1}^{N}\sum_{J=1}^{M}  \frac{Z_{J}}{r_{iJ}} + \sum_{i=1}^{N}\sum_{j>i}^{N}  \frac{1}{r_{ij}},
\end{equation}
with $r_{ij}= |\bm r_i - \bm r_j|$ being the distance between any two particles (electrons or nuclei) and $Z_J$ indicating the charge of nuclei $J$.
In the above equation, the first term is the kinetic energy of the electrons, the second term describes the electron--nucleus attraction (also referred to as the external potential), while the last term corresponds to the potential energy of the repulsion between electrons.
Usually, the nucleus--nucleus repulsion term $\sum_{I<J}^{M}\frac{Z_I Z_J}{R_{IJ}}$ is included in the electronic Hamiltonian and manifests itself as a constant shift in the electronic energy.

When the speed of the electrons becomes comparable to the speed of light, relativistic effects have to be included into the equation, which has to be invariant under Lorentz transformation.
In the framework of relativistic quantum chemistry, any free particle with spin of \nicefrac{1}{2} is described by the time-independent Dirac equation \cite{relativistic_dirac} (again in atomic units)
\begin{equation} \label{eq:fpdirac}
H_{\rm fp} \psi(\boldsymbol{x}) = \Big( c \sum_{n=1}^3 \bm \alpha_n p_n + \bm \beta c^2 \Big) \psi(\boldsymbol{x}) = E \psi(\boldsymbol{x}),
\end{equation}
where $\bm \alpha_n$ and $\bm \beta$ are Dirac matrices, $c$ is the speed of light, and the wave function $\psi(\boldsymbol{x})$ is a four-component (spinor) vector.
Specifically, $\bm \alpha_n$ are written in terms of the Pauli matrices $\bm \sigma_n$ and $\bm \beta$ contains the $2\times 2$ identity matrix,
\begin{equation} \label{eq-alphabeta}
   {\pmb{\alpha}}_n = 
    \begin{pmatrix}
        0        & {\pmb{\sigma}}_n \\
        {\pmb{\sigma}}_n & 0
    \end{pmatrix}
    , \quad
    {\pmb{\beta}} = 
    \begin{pmatrix}
        {\pmb{I}}_2 & 0 \\
        0 & -{\pmb{I}}_2
    \end{pmatrix}.
\end{equation}
For atoms and molecules, the relativistic Hamiltonian can be written as a sum of one- and two-electron operators, similar to non-relativistic theory.
The one-electron part is the sum of the one-electron Dirac Hamiltonian ${H}_D$ for all electrons in the quantum system.
Specifically for the hydrogen atom (as for all one-electron systems) the Dirac Hamiltonian ${H}_D$ can be written in closed form and reads 
\begin{equation}\label{eq:dirac}
    {H}_D= {\pmb{\beta}} c^2 + c {\pmb{\alpha}} \cdot {\pmb{p}} + {V},
\end{equation}
where ${V}$ is the Coulomb potential (electron-nuclear interaction).
Although the Dirac equation is rigorous only for one-electron systems, it provides a starting point for further routines to introduce relativistic effects for molecular systems.

\subsection{Introducing Relativistic Effects}

In actinide chemistry, the most important relativistic effects include the so-called scalar relativistic effects and spin-orbit coupling.
Specifically, scalar relativistic effects are responsible for the contraction of $s$ and $p$ orbitals and the expansion of $d$ and $f$ orbitals compared with the non-relativistic Schr\"odinger equation.  
Spin-orbit coupling originates from interactions between the magnetic field produced by the orbital motion of a charged particle and its spin.
Both scalar relativistic and spin-orbit effects are important in actinide compounds, while other higher-order effects are typically neglected~\cite{Tecmer2014}.

The most rigorous procedure to include relativistic effects is to find the eigenfunctions and eigenvalues of the four-component Dirac equation in an all-electron basis.
This many-particle equation is built on a top of the Dirac equation for a single fermion.
Specifically, the many-electron relativistic Hamiltonian combines the one-electron Dirac operators from eq.~\eqref{eq:dirac}, the electron--electron repulsion term as given in eq.~\eqref{eq:schroedingerH}, and the Breit operator \cite{relativistic_breit},
\begin{equation}\label{eq:breit}
 {g}_{ij}^{\text{Breit}} =  -\frac{c\pmb{\alpha}_i \cdot c\pmb{\alpha}_j}{2c^2r_{ij}}-\frac{(c\pmb{\alpha}_i {\bf r}_{ij}) \cdot (c\pmb{\alpha}_j {\bf r}_{ij})}{2c^2r^3_{ij}},
\end{equation}
(or the simplified Gaunt operator ${g}_{ij}^{\text{Gaunt}} =  -\frac{c\pmb{\alpha}_i \cdot c\pmb{\alpha}_j}{c^2r_{ij}}$ \cite{relativistic_gaunt}) that mimics the retardation of the potentials due to the finite speed of light.
Although the corresponding equation is Lorentz-invariant only approximately, it describes relativistic effects most accurately.
The drawback of the so-called Dirac--Coulomb--Breit Hamiltonian is the large computational cost, which makes this approach computationally infeasible for molecules with a large number of electrons. 
In practical applications the four-component Dirac--Coulomb Hamiltonian is used at the SCF level and the correlated calculations are performed within the so-called ``no-pair" approximation, where projection operators remove any Slater determinant containing negative-energy orbitals from the Dirac--Coulomb Hamiltonian~\cite{Trond_rev_2012}. 
In this approach both one- and two-electron contributions to spin--orbit coupling are accounted for. It is possible to further reduce the computational cost and approximate the ``full'' spin--orbit operator using either atomic or molecular mean field theories.~\cite{relativistic_sikkema} 

Computationally less expensive methods work within a two-component framework, where the small component of the Dirac equation is eliminated.
However, this decoupling is not straightforward for many-electron systems and a number of routines have been developed during the past decades to transform the four-component form of the many-particle Dirac equation into an equation with at most two components. \cite{theochem-two-comp-methods,Tecmer2016}
One popular approach includes the so-called regular approximations. The four-component state vector is divided into a large-component spinor $\psi^L({\bm r})$ and a small-component spinor $\psi^S({\bm r})$. \cite{relativistic-zora,relativistic_chang} The atomic balance relation between these two spinors,
\begin{equation}
   \psi^S(\pmb{r}) = \left( 1+ \frac{E-{V}}{2c^2} \right)^{-1} \frac{\pmb{\sigma} \cdot \pmb{p}}{2c} \psi^L(\pmb{r}),
\end{equation}
allows us to eliminate the small component from the Dirac equation and solve the Dirac equation for the large component only, which represents a two-component equation.
The most simple flavour of the regular approximation is the zeroth order regular approximation (ZORA), where the ZORA Hamiltonian for the large component reads
\begin{equation}
     H^{\rm ZORA} = \frac{1}{2} (\pmb{\sigma} \cdot \pmb{p}) \left( 1 - \frac{{V}}{2c^2} \right)^{-1} (\pmb{\sigma} \cdot \pmb{p}) + {V}.
\end{equation}
The above (truncated) Hamiltonian includes parts of the Darwin term and all spin-orbit interactions arising from the nuclei.
However, the ZORA Hamiltonian is not gauge invariant. This deficiency can be fixed by appropriate scaling procedures or inclusion of higher order approximations. \cite{esqc2017-3-rel,theochem-two-comp-methods}


A different family of approaches aims at decoupling the electronic and positronic solutions of the Dirac Hamiltonian using a unitary transformation $U$, which makes the Dirac Hamiltonian $H_{D}$ block-diagonal with respect to the large ($h_+$) and small component ($h_-$),
\begin{equation}\label{eq-hbar}
\bar{{H}}_{D} = {U}^\dagger {H}_D {U} =  
     \begin{pmatrix}
        {h}_+ & 0 \\
        0   & {h}_-
    \end{pmatrix}.
\end{equation}
The resulting blocks in the transformed Hamiltonian $\bar{{H}}_{D}$ are two-component Hamiltonians and act on electronic and positronic states only.
The exact form of the unitary transformation $U$ is, however, only known for the free-particle Dirac equation and is called the Foldy--Wouthuysen transformation. \cite{dhk-foldy}
An approximate decoupling scheme for the many-electron Dirac equation in quantum chemistry was proposed by Hess.
The so-called Douglas--Kroll--Hess (DKH) method~\cite{dkh1,dkh2,Reiher2006} is based on the Foldy--Wouthuysen transformation~\cite{dhk-foldy} and represents an order-by-order expansion (in the external potential $V$), where the electronic and positronic components of the Dirac equation are separated iteratively.
The DKH transformed Hamiltonian of $(n+1)$-th order has the general form
\begin{equation}
    {H}_{n+1} =U_n^\dagger U_{n-1}^\dagger\ldots U_2^\dagger U_1^\dagger {H}_1 U_1 U_2\ldots U_{n-1} U_n,
\end{equation}
where $H_1$ is the free-particle Foldy--Wouthuysen (fpFW) transformed Dirac Hamiltonian $H_1 = {U_{\rm fpFW}}^\dagger {H}_D {U_{\rm fpFW}}$.
Thus, different orders of approximations are obtained by applying subsequent unitary transformations to the relativistic Dirac Hamiltonian.~\cite{dhk2-wolf,Reiher2006,dhk2-wullen,Reiher2012}
Specifically, the second-order Douglas--Kroll--Hess (DKH2) Hamiltonian is most commonly used in quantum chemistry as it provides satisfactory results for conventional chemical problems.
In DKH2, only one unitary transformation $U_1$ has to be applied.
We should note that the explicit form of the unitary transformation $U$ does not affect lower order DKH Hamiltonians and hence the operators $U_i$ can be represented in different ways, using, for instance, a power series expansion of an (anti-Hermitian) operator.

The exact two-component (X2C) relativistic Hamiltonian is based on exact decoupling of the large and small components of the Dirac Hamiltonian in its matrix representation.
Specifically, the X2C method exploits the non-symmetric Algebraic Riccati Equation (nARE) \cite{Ricatti-proceedings,Ricatti}, a quadratic matrix equation. 
The nARE approach was used for the Dirac Hamiltonian for the first time by Ilias and Saue \cite{relativistic_x2c} and introduced as the X2C method. Most importantly, the eigenvalues of the X2C Hamiltonian are identical to the positive energy branch of the four-component Dirac Hamiltonian. 

One should stress that in the majority of quantum chemical applications, these two-component Hamiltonians account only for scalar relativistic effects and thus only have a one-component form.
Due to this one-component nature, such Hamiltonians can be easily interfaced with standard quantum chemistry codes. 
Spin-orbit coupling effects can be included \textit{a posteriori} using the spin-orbit configuration-interaction approach, where the relativistic Hamiltonian is diagonalized in the spin-free basis.~\cite{rassi,relativistic_2} To further decrease the computational cost, the spin-orbit integrals are often calculated within the atomic mean-field intergrals (AMFI) approach.~\cite{amfi_1,amfi_2,amfi_3}

The computationally most efficient way of including relativistic effects in the Schr\"odinger equation is to introduce scalar relativistic effects using relativistic effective core potentials (RECP)~\cite{recp_1}.
Such a crude approximation is usually sufficiently accurate for chemistry as the influence of the core electrons on the valence shell (that is the shell containing electrons of relevance in chemical processes) is rather indirect and can be accurately modelled using parametrized effective pseudo-potentials in conjunction with scalar relativistic interactions.~\cite{recp_1}
Besides being computationally inexpensive and fast to compute, RECPs provide reliable results if spin-orbit coupling is negligible. 
Spin--orbit corrections can be added \textit{a posteriori} on top of RECP.~\cite{recp_1,relativistic_3}

\subsection{Solving the Electronic Problem}

Since the Schr\"{o}dinger or Dirac equation cannot be solved exactly for many-electron systems, many approximate methods have been introduced to quantum chemistry that aim at solving the electronic problem as accurately as possible.
The simplest---and probably the most important---model is the molecular orbital approximation, where each electron occupies exactly one orbital.
The total electronic wave function is then constructed as an antisymmetric product of these spin orbitals $\chi_i(\bm x_j)$ that depend on the spatial coordinates $\bm r_j$ and spin coordinate $\sigma_j$ of one electron.
The antisymmetric product of spin orbitals is called a Slater determinant (or electronic configuration),
\begin{equation} \label{slater}
\Psi_{\rm el}(\boldsymbol{x}_1, \boldsymbol{x}_2, \dots, \boldsymbol{x}_N) = \frac{1}{\sqrt{N!}}\left|
\begin{matrix}
\chi_{1}(\boldsymbol{x}_{1}) & \chi_{2}(\boldsymbol{x}_{1})  & \dots & \chi_{N}(\boldsymbol{x}_{1})  \\ 
\chi_{1}(\boldsymbol{x}_{2}) & \chi_{2}(\boldsymbol{x}_{2}) & \dots & \chi_{N}(\boldsymbol{x}_{2}) \\
\vdots & \vdots &\ddots &\vdots \\
\chi_{1}(\boldsymbol{x}_{N}) & \chi_{2}(\boldsymbol{x}_{N}) & \dots & \chi_{N}(\boldsymbol{x}_{N}) 
\end{matrix}
\right|,
\end{equation} 	
where $\chi_{i}(\boldsymbol{x}_j)$ is the $i$th spin orbital populated by the $j$th electron and $N$ is the total number of electrons.
In quantum chemistry, the Hartree--Fock method optimizes a single Slater determinant and represents a common starting point for more elaborated approaches.
Using the notation of second quantization ~\cite{Helgaker_book}, a Slater determinant can be written in a very compact form,
\begin{equation} \label{slater-sq}
\Psi_{\rm el} = \prod_i a^\dagger_i | \rangle,
\end{equation}
where $a^\dagger_i$ is the fermionic creation operator, which creates an electron in spin orbital $i$, and $ | \rangle$ is the vacuum state.
For simplicity, we have dropped the dependence of $\Psi_{\rm el}$ on the electronic coordinates.
Note that a Slater determinant contains only occupied orbitals. If the number of one-electron functions (that is, spin orbitals) is greater than the total number of electrons in the system, it is possible to construct more than one Slater determinant.
If the electronic wave function is expanded as a sum of all possible Slater determinants $\Phi_k$ that can be constructed by distributing $N$ electrons in $K$ orbitals,
\begin{equation}\label{eq:fci}
\Psi_{\rm el}^{\rm FCI} = \sum_k c_k \Phi_k = \sum_k c_k \left(\prod_{i_k} a^\dagger_{i_k} | \rangle \right),
\end{equation}
we obtain the so-called Full Configuration Interaction expansion (for a given finite basis with $K$ orbitals), where $ c_k $ are some expansion coefficients.

The energy difference between the FCI solution and the electronic energy corresponding to a single Slater determinant (SD),
\begin{equation}
E_{\rm el}^{\rm corr} = E_{\rm el}^{\rm FCI}-E_{\rm el}^{\rm SD},
\end{equation}
is defined as the correlation energy and originates from the correlated motions of the electrons that cannot be described within Hartree--Fock theory (except of exchange correlation).
Thus, in order to account for correlation effects, we have to include more than one Slater determinant in the wave function expansion.
Although FCI allows us to solve the Schr\"odinger (or Dirac) equation exactly (within a given finite orbital basis), it is computationally feasible only for the smallest systems, containing up to, say, 20 electrons.
Since actinide atoms and actinide-containing molecules usually contain much more than 20 electrons, the FCI ansatz cannot be applied in computational actinide chemistry.
Furthermore, since electron correlation effects are crucial for a reliable description of chemical properties and chemical reactions involving actinide compounds, we have to find suitable wave function models that allow us to approximate the FCI wave function as accurate as possible by reducing the number of degrees of freedom in the optimization problem.
This can be done by either restricting the number of Slater determinants by truncating the FCI expansion or by using more efficient parameterizations of the CI expansion coefficients $c_k$ (or any combinations of those two strategies).

\subsubsection{Accounting for Electron Correlation Effects in the Ground-state Electronic Wave Function}

In quantum chemistry, we usually distinguish between single- and multi-reference approaches.
The former employ some reference configuration $\Phi_0$ to construct a truncated CI expansion. Multi-reference methods do not refer to a single Slater determinant but employ a set of selected determinants that are chosen due to some criterion.
Both single- and multi-reference methods are commonly applied in computational actinide chemistry to model ground- and excited-states properties.
In the following, we will briefly discuss some conventional and unconventional electronic structure methods that have been used to study heavy-element-containing compounds.

\subsubsection{Truncated Configuration Interaction}

One single-reference approach, where the FCI wave function is systematically truncated, represents truncated configuration interaction (CI).
In truncated CI, only those Slater determinants are included in the wave function expansion that differ by one, two, three, etc. orbitals with respect to the reference determinant $\Phi_0$.
The electronic wave function is then a linear expansion containing the reference determinant and all singly, doubly, triply, etc. substituted configurations.
Most commonly, the FCI expansion is truncated to include only single and double excitations leading to the CI Singles Doubles (CISD) wave function,
\begin{equation}
    \Psi_{\rm el}^{\rm CISD} = \Phi_0 + \sum_i^{\rm occ} \sum_a^{\rm virt} c_i^a a^\dagger_{a} a_i \Phi_0
                                      + \sum_{i<j}^{\rm occ} \sum_{a<b}^{\rm virt} c_{ij}^{ab} a^\dagger_{a} a^\dagger_b a_j a_i \Phi_0.
\end{equation}
In the above equation, we have used the conventional notation of quantum chemistry, where indices $i,j,\ldots$ indicate occupied (spin) orbitals, while $a,b,\ldots$ run over all virtual (spin) orbitals of the reference determinant $\Phi_0$.
$a_i$ is the fermionic annihilation operator and depopulates the $i$-th orbital.
One drawback of CISD (or any truncated CI method) is its lack of size-extensivity and size-consistency.
The size-consistency error can be reduced using, for instance, the Davidson correction.~\cite{davidson-correction}

\subsubsection{Single-reference Coupled Cluster Theory}

A different single-reference method that is frequently applied in actinide chemistry is coupled cluster (CC) theory.
In the CC method, the electronic wave function is written using an exponential ansatz,
\begin{equation}
    \Psi_{\rm el}^{\rm CC} = e^{T}\Phi_0,
\end{equation}
where $T$ is the so-called cluster operator and can be expressed as a sum of excitation operators $T = T_1+ T_2 + T_3 + \ldots$.
As in truncated CI, the excitation operators excite one, two, three, etc.~electrons from occupied orbitals to virtual orbitals,
\begin{equation}
    T_1 = \sum_i\sum_a t_i^a a^\dagger_a a_i, \quad
    T_2 = \frac{1}{(2!)^2}\sum_{ij}\sum_{ab} t_{ij}^{ab} a^\dagger_a a^\dagger_b a_j a_i,\quad
    T_3 = \frac{1}{(3!)^2}\sum_{ijk}\sum_{abc} t_{ijk}^{abc} a^\dagger_a a^\dagger_b a^\dagger_c a_k a_j a_i,
\end{equation}
and so on, where $t_i^a, t_{ij}^{ab}, \ldots$ are the CC singles, doubles, etc.~amplitudes.
In conventional electronic structure calculations, the full cluster operator is approximated and restricted to include only some lower-order excitation operators.
Specifically, in the CC Singles and Doubles (CCSD) approach, we have $T = T_1+T_2$.
In truncated CC methods, the wave function expansion still contains all Slater determinants of the FCI expansion, yet the expansion coefficients $c_k$ are approximated by only a subset of cluster amplitudes.
These conventional single-reference methods typically break down when orbitals become (quasi-)degenerate and hence cannot be unambiguously separated into an occupied and virtual space.
In such strongly-correlated cases, we can switch to a multi-reference description of electronic structures.
There exist, however, extensions (or simplifications) of conventional single-reference methods that allow us to model strongly-correlated quantum states within a single-reference framework.
Examples are spin-flip CC~\cite{spin-flip-EOM} and pair-CCD~\cite{geminals_6}. 

\subsubsection{Multi-reference Complete Active Space Self-consistent Field Theory}

The complete active space self-consistent field (CASSCF) method is a variant of the multi-configurational SCF (MCSCF) approach.
The model wave function,
\begin{equation}
{\Psi_{\rm el}^{\rm CASSCF}} = e^{-{\kappa}} \sum_i c_i {\Phi_i},
\end{equation}
has a similar form as the FCI wave function being a linear expansion in terms of Slater determinants (or configuration state functions) ${\Phi_i}$ with expansion coefficients $c_i$.
The operator $e^{-{\kappa}}$ performs unitary transformation of the spin orbitals, where ${\kappa}$ is the generator of orbital rotations,
\begin{equation}\label{eq:kappa}
{\kappa} = \sum_{p>q} \kappa_{pq} (a^\dagger_p a_q - a^\dagger_q a_p).
\end{equation}
Thus, in contrast to FCI, the orbital basis is optimized self-consistently within CASSCF.
The ground-state wave function is obtained by minimizing the electronic energy with respect to all variational parameters,
\begin{equation}
E_{\rm el}^{\rm CASSCF} = \min_{\bm \kappa, \bm c} \frac{\bra{\Psi_{\rm el}^{\rm CASSCF}(\bm \kappa, \bm c)} H \ket{\Psi_{\rm el}^{\rm CASSCF}(\bm \kappa, \bm c)}}{\bra{\Psi_{\rm el}^{\rm CASSCF}(\bm \kappa, \bm c)}\ket{\Psi_{\rm el}^{\rm CASSCF}(\bm \kappa, \bm c)}},
\end{equation}
So far, we have made no assumptions about the configurational space of CASSCF and the above equations are valid for any MCSCF wave function.
Since we have to optimize both the expansion coefficients $c_i$ and the spin orbitals, MCSCF-type methods are computationally expensive.
To reduce the computational cost, the configurational space is heavily truncated.
Specifically in CASSCF, the molecular orbitals are divided into three subsets: (1) doubly-occupied inactive (frozen, core) orbitals, (2) active orbitals, and (3) unoccupied external (virtual) orbitals.
In each electronic configuration (Slater determinant), the inactive orbitals are always doubly occupied, while the external orbitals remain unoccupied.
Only the orbital occupations of the active orbitals are allowed to differ in each Slater determinant.
Furthermore, all possible ways of distributing the active electrons in the active space orbitals are permitted in the CASSCF wave function, which makes the active space complete in terms of the CI expansion.
Thus, CASSCF represents a FCI expansion in the active space orbitals.

In general, the active space should compromise all chemically important orbitals for a given molecular system.
For small molecules, an energetic criterion can be used to select the active space orbitals.
In actinide chemistry, conventional selection procedures might be ineffective as actinide complexes feature many quasi-degenerate orbitals and it remains ambiguous which metal and ligand orbitals have to be included in the active space.
Novel approaches based on quantum information theory allow us to identify the chemically most important orbitals and can be applied to develop a black-box like selection procedure of active spaces for MCSCF-type calculations.
Such an approach will be discussed in section \ref{sec:qit}.
The CASSCF method removes the problem of (quasi-)degeneracies and allows us to model static correlation effects.
However, it does not account for dynamical correlation effects attributed to electron excitations beyond the active space orbitals.
CASSCF, thus, provides a spin-adapted zero-order wave function, where dynamical and core-valence electron correlation can be added using various \textit{a posteriori} corrections such as complete-active-space second-order perturbation theory (CASPT2)~\cite{caspt21,caspt22} or multi-reference configuration-interaction (MRCI)~\cite{mrci}.

\subsubsection{The Density MatrixRenormalization Group}

The density matrix renormalization group (DMRG)~\cite{dmrg-book-1,dmrg-4,dmrg-5,dmrg-6,dmrg-7,dmrg-8,dmrg-9,dmrg-10,dmrg-21} algorithm represents a computationally efficient variant of MCSCF theory where the evaluation of the electronic energy scales only polynomially with system size.
Due to its low computational scaling, the DMRG protocol allows us to approach the FCI limit of an $N$-particle Hilbert space constructed from $L$ orbitals for large molecules, where FCI calculations are computationally unfeasible.
In contrast to conventional \textit{ab initio} methods, DMRG optimizes a special type of quantum states, so-called matrix product states (MPS), that allow us to efficiently reparameterize the electronic wave function using a significantly smaller number of variational parameters.
In the MPS representation, the CI wave function expansion eq.~\eqref{eq:fci} is rewritten in terms of a product of matrices that replaces the CI expansion coefficients,
\begin{equation}
\Psi_{\rm el}^{\rm DMRG} = \sum_{k_1, k_2, \dots, k_L} \bm A_1^{k_1} \bm A_2^{k_2} \dots {\bm A}_{L}^{k_L} |k_{1}, k_{2}, \dots, k_{L} \rangle,
\end{equation}
where $L$ is the number of spatial orbitals in some active space, $\{\bm A_1^{k_1}, \bm A_2^{k_2} ,\dots, \bm A_{L}^{k_L}\}$ is a set of matrices that are optimized by the algorithm, and $\{k_{1}, k_{2}, \dots, k_{L}\}$ are the occupations of the orbitals (either unoccupied, singly occupied, or doubly occupied) written in terms of the occupation number representation, where each occupation number vector $\ket{k_{1}, k_{2}, \dots, k_{L}}$ represents a Slater determinant.

The DMRG wave function and its many-particle basis is optimized in a sweeping procedure.
One sweep contains $(L-q-2)$ microiterations, where $q$ is the number of exactly-represented orbitals (either 1 or 2).
To perform the sweeping algorithm, the orbitals have to be aligned on a one-dimensional lattice.
Thus, the DMRG algorithm is best suited to describe one-dimensional problems.
There exist different approaches to order the orbitals along a one-dimensional lattice.
Specifically, concepts of quantum information theory allow us to select an optimal orbital ordering in a black-box-like fashion.~\cite{DEAS}
One microiteration includes three distinct steps: (1) blocking, (2) diagonalization, and (3) decimation.
First, the $L$-orbital space is partitioned into three subspaces: a system block, an environment block, and $q$ exactly represented orbitals in between represented by 4 basis functions in the case of spatial orbitals.
In the blocking step, the system and, if $q=2$, the environment are enlarged by one of the exactly represented orbitals.
The many-particle basis states are defined as a tensor product of basis states of the subsystem block and the neighbouring exactly-represented orbital.
In the second step, the Hamiltonian of the \textit{superblock} (enlarged system+enlarged environment block) is constructed and diagonalized.
Usually, we are only interested in the ground-state wave function and hence only one root of the superblock Hamiltonian needs to be computed.
In the third and last step, the dimensionality of the enlarged system and enlarged environment blocks is reduced to prevent the many-particle basis from growing exponentially (due to the blocking step).
The number of basis functions is reduced to a limit indicated by a parameter $m$.
In DMRG, this so-called renormalization step is performed in a specific way.
From the superblock wave function $\Psi^{\rm SB}$, we calculate the many-particle reduced density matrix of the enlarged active system block $\rho^{\rm s}= \Tr_{m_{\rm e}} \ket{\Psi^{\rm SB}}\bra{\Psi^{\rm SB}}$, where $m_{\rm e}$ indicate states defined on the (enlarged) environment block.
This reduced density matrix $\rho^{\rm s}$ is then diagonalized and the eigenvectors corresponding to the $m$ largest eigenvalues form the new many-particle basis of the enlarged system block.
In the final renormalization step, all matrix representations of operators are transformed into this new basis.
Specifically, the computed transformation matrices correspond to the $\bm A$ matrices of the MPS ansatz.
Thus, each microiteration step optimizes exactly one MPS matrix $\bm A$ and we have to sweep through the lattice to obtain an (approximate) full representation of the MPS.
After the decimation step, the algorithm starts again with the blocking procedure, where the new system block is enlarged by one orbital, while the new environment is reduced by one orbital.
We should note that the choice of $m$ is crucial to find a balance between accuracy and computational cost.
There is, however, no straightforward formula that indicates the best value of $m$ and several calculations with different values of $m$ are required to converge the wave function with satisfactory accuracy.

Typically, an MPS is represented in its canonical form containing so-called left- and right-normalized matrices.
The DMRG algorithm optimizes a mixed-canonical MPS that is composed of both left- and right-normalized matrices,
\begin{equation}\label{eq:mixedmps}
\Psi^{\rm MPS}_{\rm el}  = \sum_{k_1, k_2, \dots, k_L} \bm A_1^{k_1} \bm A_2^{k_2} \dots {\bm A}_{l-1}^{k_{l-1}} {\bm \Psi}^{k_{l} k_{l+1}} \bm A_{l+2}^{k_{l+2}} {\dots}  \bm A_{L}^{k_L} |k_{1}, k_{2}, \dots, k_{L} \rangle.
\end{equation}
In the above equation, the left-normalized matrices are defined for orbitals $k_1,\ldots,k_{l-1}$, while the right-normalized matrices are obtained for orbitals $k_{l+2},\ldots,k_{L}$.
The matrix ${\bm \Psi}^{k_{l} k_{l+1}}$ is obtained during the diagonalization step of the superblock Hamiltonian.
Since one matrix of the MPS representation contains at most $m^2$ elements, the total number of variational parameters is at most $4Lm^2$ if we work in a spatial orbital basis with 4 possible occupations.
Thus, the high-dimensional CI coefficient tensor, which scales binomially with system size, has been replaced by a product of lower-dimensional tensors. 

The DMRG algorithm is a powerful tool to approximate FCI wave functions in a given active space that are computationally not accessible for conventional quantum mechanical methods like CASSCF.
Most importantly, it is suitable for strongly-correlated systems and hence allows us to accurately model heavy-element compounds, like transition metal- or actinide-containing molecules.
Although originally formulated to tackle one-dimensional problems, DMRG has been successfully applied to describe strong correlation in general 3-dimensional systems.
The missing dynamical correlation effects can be added \textit{a posteriori} using the same corrections as developed for traditional MCSCF methods.
Examples are second order complete active space perturbation theory (DMRG-CASPT2),~\cite{dmrg-caspt2} the multi-reference configuration interaction approach with internal contraction of DMRG (DMRG-icMRCI),~\cite{dmrg-icmrci} canonical transformation theory (CT),~\cite{canonical-transformation} or perturbation theory formulated in terms of matrix product states.~\cite{dmrg-mrpt-mps}

\subsubsection{Geminal-based Approaches}

All electronic structure methods discussed above use one-electron functions (orbitals) to construct the Slater determinants that span the $N$-particle Hilbert space.
A conceptionally different approach to account for electron correlation effects is to use two-electron functions as fundamental building blocks of the electronic wave function.~\cite{geminals_7,geminals_8,geminals_9}
In second quantization, a (singlet) two-electron operator $\psi_i^{\dagger}$, also called geminal, can be written as a linear combinations of electron-pair creators,
    \begin{equation}\label{eq:geminal}
             \psi_i^{\dagger}=\sum_{q=1}^{M}C_q^{i}a_{q}^{\dagger}a_{\bar{q}}^{\dagger},
    \end{equation}
where $a_{q}^{\dagger}$ ($a_{\bar{q}}^{\dagger}$) are the fermionic creation operators for $\alpha$ ($\beta$) electrons and $M$ is the number of one-electron functions used to construct geminal $i$.
In the above equation, $C_q^{i}$ are the geminal coefficients that link the geminal creation operator with the underlying one-particle basis (represented by $a_p^\dagger$).
Thus, geminals are quasi-particles and the corresponding geminal creation operators are electron pair creators.
The geminal-based electronic wave function is a product of the geminal creation operators acting on the vacuum state,
    \begin{equation}\label{eq:geminalwfn}
             \Psi_{\rm el} = \prod_i^P \psi_i^{\dagger} \ket{},
    \end{equation}
with $P=\nicefrac{N}{2}$ being the number of electron pairs.
If the geminal creation operators have the general form of eq.~\eqref{eq:geminal}, we obtain the antisymmetric product of interacting geminals (APIG) wave function~\cite{geminals_8}.
Although APIG includes correlations between orbital pairs, it is computationally intractable for larger systems.
Substituting eq.~\eqref{eq:geminal} in the electronic wave function eq.~\eqref{eq:geminalwfn}, we can rewrite the APIG wave function using a linear expansion of Slater determinants,
    \begin{equation}\label{eq:apigwfn}
        \Psi_{\rm el}^{\rm APIG} =\sum_{\{ m_i={0,1}|P\}}|{\bm C(\bm m)}|^{+}(a_1^{\dagger}a_{\bar {1}}^{\dagger})^{m_1}(a_2^{\dagger}a_{\bar {2}}^{\dagger})^{m_2}\ldots (a_K^{\dagger}a_{\bar {K}}^{\dagger})^{m_M} \ket{},
    \end{equation}
where ${\bm C(\bm m)}$ is the geminal coefficient matrix, $|\bm A|^+$ indicates the permanent of matrix $\bm A$, and
\begin{equation}
    P=\sum_{k=1}^M m_k \,\quad {\rm with}\, \quad P < M.
\end{equation}
Specifically, ${\bm C(\bm m)}$ is a $P\times P$ matrix and contains only those columns for which $m_k=1$.
In order to evaluate the coefficients in front of the wave function expansion of eq.~\eqref{eq:apigwfn}, we have to evaluate the permanent of ${\bm C(\bm m)}$ (similar to the determinant with all $-$ signs replaced by $+$ signs).
Since the evaluation of the permanent of a matrix is $\#$P-hard and the Slater determinant expansion of eq.~\eqref{eq:apigwfn} includes a factorial number of determinants, the APIG model is computationally expensive.
To make geminal-based models applicable to larger systems, we have to introduce constraints that allow us to evaluate the permanent efficiently.
One simplified geminal-based wave function is the antisymmetric product of strongly orthogonal geminals (APSG)~\cite{geminal-apsg-1,geminal-apsg-2,geminal-apsg-3,geminal-apsg-4,geminal-apsg-5}, where the geminal creation operators create two-electron states that are orthogonal to each other.
Specifically, the sum of eq.~\eqref{eq:geminal} is restricted to run over mutually exclusive subspaces $M_i$ of orbitals,
          \begin{equation*}
              \psi_i^{\dagger}=\sum_{q=1}^{M_i}C_q^{ i}a_{q}^{\dagger i}a_{\bar{q}}^{\dagger i}\,\quad{\rm with} \,\quad \sum_qC_q^iC_q^k=\delta_{ik}.
          \end{equation*}
The partitioning of $M$ into disjoint subspaces $M_i$ is equivalent to associating subsets of orbitals to specific geminals, that is, each orbital may belong to only one geminal.
Although the strong orthogonality constraint allows us to efficiently optimize the wave function using the variational principle, we miss electron correlation effects between the orbital subsets (as they are disjoint).

A promising geminal-based model that has been successfully applied to actinide chemistry is the antisymmetric product of 1-reference orbital geminal (AP1roG)~\cite{geminals_1,geminals_9,geminals_2,geminals_3,garza2015actinide}.
In AP1roG, the strong orthogonality constraint is relaxed and inter-geminal correlations are introduced in the geminal ansatz,
        \begin{equation}\label{eq:ap1rog}
            \psi_i^\dagger = a^\dagger_i a^\dagger_{\bar{i}} + \sum_{a}^{virt} c_i^a a_a^{\dagger}a_{\bar {a}}^{\dagger},
        \end{equation}
where the sum runs over all virtual orbitals with respect to some reference determinant (like the HF determinant).
The second term of the above equations assigns (virtual) orbitals to all geminals and accounts for inter-geminal correlations.
The main feature of the AP1roG wave function is that the first term of eq.~\eqref{eq:ap1rog} selects some reference determinant, that is, one specific orbital is occupied by an $\alpha$ and $\beta$ electron in one specific geminal.
The corresponding geminal coefficient matrix has a special form,
    \begin{equation}
        {\bm C}_{ \rm AP1roG}=
        \begin{pmatrix}
  1      & \cdots  & 0      & 0 & c_{1;P+1} & c_{1;P+2}&\cdots &c_{1;K}\\
  0      & 1       & \cdots & 0 & c_{2;P+1} & c_{2;P+2}&\cdots &c_{2;K}\\
  \vdots & \vdots  & \ddots & \vdots  & \vdots    & \vdots   &\ddots &\vdots\\
  0      & \cdots  & 0 & 1 & c_{P;P+1} & c_{P;P+2}&\cdots & c_{P;K}
 \end{pmatrix},
    \end{equation}
where each row is one geminal and the left block contains the $P\times P$ identity matrix due to the first term in eq.~\eqref{eq:ap1rog}.

In addition to the above mentioned geminal models, different geminal-based wave functions have been introduced in quantum chemistry, like generalized-valence-bond perfect-pairing (GVB-PP)~\cite{geminal-gvb-1,geminal-gvb-2,geminal-gvb-3} and the particle-number projected Hartree--Fock--Bogoliubov model~\cite{geminal-hfg}.
However, none of these geminal-based models have been applied to actinide chemistry and hence will not be discussed in this chapter.
In the following, we will have a closer look at the AP1roG method, its optimization schemes, and possible extensions, as it has been proven to properly describe the static correlation in certain (heavy-element containing) molecules such as \ce{UO2^{2+} and \ce{ThO_2}}.~\cite{uranyl-dissociation}

Although the structure of the AP1roG coefficient matrix allows us to efficiently evaluate the permanent $|\bm C|^+$, we still have to deal with a factorial number of Slater determinants when optimizing the electronic wave function.
To obtain a computationally efficient optimization method, we can rewrite the AP1roG wave function using an exponential ansatz,
\begin{align}
    {\Psi_{\rm el}^{\rm AP1roG}} &= \prod_i \psi_i^\dagger \ket{} \nonumber \\
                                 &= \prod_i \left ( a^\dagger_i a^\dagger_{\bar{i}} + \sum_{a}^{virt} c_i^a a_a^{\dagger}a_{\bar {a}}^{\dagger} \right ) \ket{} \nonumber \\
                                 &= \prod_i \left ( 1 + \sum_{a}^{virt} c_i^a a_a^{\dagger}a_{\bar {a}}^{\dagger} a_{\bar{i}} a_i  \right )  a^\dagger_i a^\dagger_{\bar{i}} \ket{} \nonumber \\
                                 &= e^{\sum_i^{\rm occ}\sum_{a}^{virt} c_i^a a_a^{\dagger}a_{\bar {a}}^{\dagger} a_{\bar{i}} a_i } \Phi_0,
\end{align}
where $\Phi_0 = \prod_i a^\dagger_i a^\dagger_{\bar{i}} \ket{}$.
Thus, the AP1roG method optimizes a coupled cluster-type wave function where the cluster operator $T$ is restricted to electron pair excitations $T_\textrm{p}$~\cite{geminals_6},
\begin{equation}
    {T}_\textrm{p} = \sum_{i}^{\rm occ} \sum_{a}^{\rm virt} c_{i}^a a_{a}^{\dagger} a_{\bar{a}}^{\dagger} a_{\bar{i}} a_{i}.
\end{equation}
Since $T_\textrm{p}$ excites an electron pair, it can be considered as a simplified $T_2$ operator and hence AP1roG is a simplified version of the CCD method.
Due to its exponential ansatz, the AP1roG wave function is also known as the pair coupled cluster doubles (pCCD) wave function,
\begin{equation}\label{eq:ap1rog-pccd}
{\Psi_{\rm el}^{\rm AP1roG}} ={\Psi_{\rm el}^{\rm pCCD}}= e^{{T}_\textrm{p}} \Phi_0.
\end{equation}
Furthermore, the geminal coefficients $\{c_i^a\}$ of AP1roG are equivalent to the pCCD amplitudes and we can use the optimization techniques of single-reference coupled cluster theory to solve for $\{c_i^a\}$.
Specifically, the electron-pair amplitudes are optimized using the projected Schr\"odinger equation, where the projection manifold is restricted to electron-pair excited determinants $\Phi_{i\bar{i}}^{a\bar a} = a_a^\dagger a_{\bar{a}}^{\dagger} a_{\bar{i}} a_{i} \Phi_0$,~\cite{geminals_7}
\begin{equation}\label{eq:ap1rogopt}
    \bra{\Phi_{i\bar{i}}^{a\bar a} } H \ket{\Psi_{\rm el}^{\rm AP1roG}} = E \bra{\Phi_{i\bar{i}}^{a\bar a} }\ket{\Psi_{\rm el}^{\rm AP1roG}} = E c_i^a,
\end{equation}
or using the similarity transformed Hamiltonian of coupled cluster theory,~\cite{geminals_6}
\begin{equation}\label{eq:pccdopt}
    \bra{\Phi_{i\bar{i}}^{a\bar a} } e^{-T_\textrm{p}}H e^{T_\textrm{p}} \ket{\Phi_0} = 0.
\end{equation}
Both optimization procedures are equivalent and eqs.~\eqref{eq:ap1rogopt} and \eqref{eq:pccdopt} yield similar working equations.

Although the AP1roG wave function does not contain determinants with unpaired electrons or so-called ``broken'' electron pairs, reducing the CCD model to electron-pair terms not only decreases the computational cost, but also provides a better description of strongly-correlated systems.
The AP1roG method is thus suitable for systems with quasi-degenerate electronic states, transition-state structures, and bond-breaking processes.
The composition of the ansatz has been validated by Bytautas \textit{et al.}~\cite{seniority_1,geminals_7}, who have shown that the correct description of strong correlation effects depends mainly on determinants with a small number of unpaired electrons. 

In contrast to CCD, the AP1roG method is sensitive to rotations among the occupied orbitals and among the virtual orbitals as the pairing schemes are not equivalent.
Thus, electronic energies and properties depend on the choice of the molecular orbital basis 
and two different molecular orbital sets can yield different results even though the reference determinant remains unaffected (note that the HF determinant is invariant under rotations of the occupied or virtual orbital space, respectively).
In order to resolve the problem of non-size-consistency, the molecular orbital basis and hence the pairing scheme need to be optimized.
The optimization of the orbital-pairing scheme allows us to obtain accurate results that almost reproduce doubly-occupied self consistent field (DOSCF) results. \cite{geminals_8}
Computational studies suggest that a variational orbital optimization protocol provides the most robust and reliable orbital optimization procedure in comparison to other investigated non-variational methods. \cite{geminals_3,geminals_9}
The optimal set of orbitals is obtained by minimizing the AP1roG energy functional subject to the constraint that the AP1roG coefficient equations eqs.~\eqref{eq:ap1rogopt} are satisfied.
The energy Lagrangian, thus, reads
\begin{equation}
    \mathcal{L} = \bra{\Phi_0} e^\kappa {H} e^{-\kappa} \ket{\Psi_{\rm el}^{\rm AP1roG}}
                    + \sum_{i,a} \lambda_i^a \big(\bra{\Phi_{i\bar i}^{a \bar a}} e^\kappa {H} e^{-\kappa} \ket{\Psi_{\rm el}^{\rm AP1roG}} - Ec_i^a \big),
\end{equation}
where $\kappa$ is again the generator of orbital rotations as defined in eq.~\eqref{eq:kappa} and \{$\lambda_i^a$\} are the Lagrange multipliers.
The Lagrange multipliers are obtained from equations that are analogous to the $\Lambda$-equations of coupled cluster theory, where we require the partial derivative of the Lagrangian with respect to the geminal coefficients \{$c_i^a$\} to equal zero, $\frac{\partial\mathcal{L}}{\partial c_i^a }\vert_{\kappa =0}=0$.
The geminal coefficients are obtained by making $\mathcal{L}$ stationary with respect to the Lagrange multipliers \{$\lambda_i^a$\}, $\frac{\partial\mathcal{L}}{\partial \lambda_i^a }\vert_{\kappa =0}=0$.~\cite{geminals_3}
The orbital gradient is the partial derivative of $\mathcal{L}$ with respect to the orbital rotation coefficients $\{\kappa_{pq}\}$ evaluated for the current set of orbitals ($\kappa = 0$),
\begin{align}\label{eq:voo-grad}
\frac{\partial \mathcal{L}}{\partial \kappa_{pq}}\Big\vert_{\kappa=0} = g_{pq}\vert_{\kappa=0} = 
               &\bra{ \Phi_0 + \sum_{i,a} \lambda_i^a \Phi_{i \bar{i}}^{a \bar{a}} } [(a^\dagger_p a_q - a^\dagger_q a_p),{H}] \ket{\Psi_{\rm el}^{\rm AP1roG}}  \nonumber \\
               &- \sum_{i,a} \bra{\Phi_0}[(a^\dagger_p a_q - a^\dagger_q a_p), {H}] \ket{\Psi_{\rm el}^{\rm AP1roG}} \sum_{i,a} \lambda_i^a c_i^a .
\end{align}
After the orbital gradient and (approximate) orbital Hessian $\bm A$ are evaluated, the matrix representation of $\kappa$ can be determined from 
\begin{equation}\label{eq:kappa}
{\bm \kappa} = -{\bm A}  {\bm g}
\end{equation}
and the orbital basis can be transformed using the unitary transformation matrix $e^{-\kappa}$.
For reasons of computational efficiency, the orbital Hessian is typically approximated by its diagonal, $\bm{A}_{pq,pq}=\frac{\partial \bm {g}_{pq}}{\partial \kappa_{pq}}\big\vert_{{\kappa}=0}$.


Although AP1roG captures a significant amount of the strong electron correlation energy and represents a very promising reference wave function in actinide chemistry, it misses a large fraction of the dynamic (weak) correlation energy.
Dynamic electron correlation effects on top of the geminal wave function can be included in the wave function ansatz \textit{a posteriori} using, for instance, perturbation theory~\cite{geminal-mrpt,geminal-pt-1,geminal-pt-2,geminal-pt-3,geminal-pt-4}, extended random phase approximation \cite{geminal-erpa1,geminal-erpa2,geminal-erpa3,geminal-erpa4}, density functional theory (DFT) \cite{geminal-erpa3, geminal-dft, geminal-dft2}, and coupled cluster theory \cite{geminals_6} or its linearized version \cite{geminal-erpa3, geminal-lcc-2, geminals_lcc_2015}.
Numerical studies indicate that the perturbation theory corrections with an AP1roG reference function do not provide reliable electronic structures and properties for actinide-containing compounds.
To reliably model actinide chemistry, we can use various coupled cluster corrections on top of the AP1roG wave function.

One possible way to extend AP1roG is to apply an AP1roG-tailored CC formalism.
In AP1roG-tailored CC theory, the electron-pair amplitudes of the CC singles, doubles, triples, etc.~equations are substituted by the AP1roG geminal coefficients and not optimized, that is kept frozen, during the optimization procedure.
Note that the tailored CC amplitude equations are similar to the conventional CC working equations, except that some selected amplitudes (the tailored amplitudes) are not varied.
Thus, a tailored CC calculation represents only a minor modification in any CC code.
In the case of AP1roG-tailored CC, the corresponding CC corrections are referred to as frozen-pair (fp) CCD, fpCCSD, fpCCSDT, etc. 
In the fpCCD and fpCCSD methods, the single and non-pair double amplitudes can provide a balanced description of electron correlation effects when both CCD and CCSD fail in describing strongly-correlated systems.

A different CC corrections on top of AP1roG employs a linearized coupled cluster (LCC) ansatz and represents a simplification of any frozen-pair CC approach.
In the LCC correction, we approximate the exponential coupled cluster ansatz with an AP1roG reference as
\begin{align}
{\Psi_{\rm el}^{\rm AP1roG-LCC}}    &= e^{{T}} \Psi_{\rm el}^{\rm AP1roG} \label{eq:ap1roglcc}\\
                                    &\approx (1+T) \Psi_{\rm el}^{\rm AP1roG},
\end{align}
where ${T}$ is a general cluster operator.
The Schr\"{o}dinger equation for this wave function ansatz reads
\begin{align}\label{eq:hlcc}
    {H} \ket{\Psi_{\rm el}^{\rm AP1roG-LCC}} &= E \ket{\Psi_{\rm el}^{\rm AP1roG-LCC}} \nonumber \\
    e^{-{T}}{H} e^{{T}} \ket{\Psi_{\rm el}^{\rm AP1roG}} &= E \ket{\Psi_{\rm el}^{\rm AP1roG}},
\end{align}
where we have used eq.~\eqref{eq:ap1roglcc} and multiplied from the left by $e^{-T}$.
In the LCC correction, the left-hand-side of eq.~\eqref{eq:hlcc} is approximated to contain only linear terms in the Baker--Campbell--Hausdorff expansion,
\begin{equation}\label{eq:bchlcc}
    ({H} + [{H}, {T}] ) \ket{\Psi_{\rm el}^{\rm AP1roG}} = E \ket{\Psi_{\rm el}^{\rm AP1roG}}.
\end{equation}
If we now substitute the exponential form of the AP1roG wave function eq.~\eqref{eq:ap1rog-pccd} in the above equation, we can bring the AP1roG-LCC Schr\"odinger equation into the familiar form
\begin{equation}\label{eq:lcceq}
    ({H} + [{H}, {T}] + [[{H}, {T}],T_{\rm p}]) \ket{\Phi_0} = E \ket{\Phi_0}
\end{equation}
of single-reference CC theory.
Furthermore, in AP1roG-LCC, the cluster operator is restricted to contain electron excitations (singles, broken-pair doubles, etc.) beyond electron-pair excitations.
For instance, in the case of double excitations, we must have $T = T_2-T_{\rm p}$, which results in the AP1roG-LCCD method.
Note that eq.~\eqref{eq:lcceq} is the Schr\"odinger equation for the non-pair amplitudes as the electron-pair amplitudes have been already optimized within AP1roG.
Although being simplifications of conventional CC methods, the linearized and frozen-pair CC corrections feature a similar computational scaling as their single-reference counter parts.

\subsubsection{Kohn-Sham Density Functional Theory}

Density functional theory (DFT) is the most popular electronic structure method due to its rather low computational cost and conceptual simplicity.
In DFT, the molecular system and its properties are determined by its electron density $\rho(\textbf{r})$ instead of the electronic wave function.
Specifically, Hohenberg and Kohn \cite{dft-hohenberg-kohn} proved that the non-degenerate ground-state wave function is uniquely determined by the electron density that corresponds to some external potential $v_{\rm ext}(\textbf{r})$.

The most common implementation of this method is within the Kohn-Sham formalism (KS) \cite{dft-ks}.
Specifically, in KS-DFT, an artificial reference system of non-interacting electrons is introduced that yields exactly the same electron density as the fully interacting system.
Furthermore, the electronic energy is a functional of the density and is decomposed into different contributions,
\begin{equation}\label{eq:dfte}
E[\rho] = T_s[\rho] + V_{\rm ext}[\rho] + J[\rho] + E_{\rm xc}[\rho],
\end{equation}
where $T_s[\rho]$ is the kinetic energy of the non-interacting system, $V_{\rm ext}[\rho]$ is the potential energy due to some external potential, $J[\rho]$ is the classical Coulomb interaction, and $E_{xc}[\rho]$ is the so-called exchange--correlation functional and accounts for all non-classical contributions to the electron--electron interaction as well as a correction term for the kinetic energy that corresponds to the difference in kinetic energy of the fully-interacting and non-interacting system.
However, the exact form of $E_{xc}[\rho]$ in eq.~\eqref{eq:dfte} is unknown and approximations thereof have to be used.
Due to their approximate nature, some density functional approximations (DFA) are appropriate for only certain types of molecules or particular properties.~\cite{dft-1,dft-2,dft-3,uran-compounds}
One major drawback of DFAs is the so-called self-interaction error attributed to the interactions between an electron and its own electric field.~\cite{SI}
This error is an artifact of the approximate nature of the DFT exchange--correlation functional.
Paradoxically, the self-interaction error may be partly balanced by other deficiencies in the energy functional yielding electronic energies and molecular properties that agree well with experimental results due to cancellation of errors.

In KS-DFT, we have to solve a set of one-particle equations (the KS equations),
\begin{equation}
\Big( -\frac{1}{2} \nabla^2 + v(\textbf{r}_i) \Big) \chi_i(\textbf{r}_i) = \epsilon_i  \chi_i(\textbf{r}_i),
\end{equation}
which optimize the KS orbitals $\chi_i(\textbf{r}_i)$.
In the above equation, $v(\textbf{r}_i)$ is an effective potential and determined as the variation of the energy functional $E[\rho]$ with respect to the electron density.
After the KS equations are solved, the electron density can be expressed in terms of the optimized KS orbitals,
\begin{equation}
\rho(\textbf{r}) = \sum_{i}^{N} |\chi_{i}(\textbf{r})|^{2}.
\end{equation}


Due to its low computational cost, KS-DFT has been extensively used in actinide chemistry, including its time dependent extensions (TD-DFT) to model electronically excited states.~\cite{dft-functional-exch,dft-functional-rev,dft_chalcogen_1,dft_chalcogen_2,dft_chalcogen_3,uran-compounds} 
Specifically, molecular structures can be accurately calculated using generalized gradient approximation (GGA) functionals, such as BP86,~\cite{perdew86,becke88}
while electronic energies (thermochemistry) and excitation energies are best determined using so-called hybrid exchange--correlation functionals, like PBE0~\cite{pbe0_1,pbe0_2,pbe0_3}, B3LYP,~\cite{becke88,b3lyp_1}, or CAM-B3LYP~\cite{CAM-B3LYP} that provide good agreement with experimental data or high-level wave-function-based methods.~\cite{dft-func-1,dft-func-2,dft-func-3,uranyl-spectra-4,actinoid_rev_2012}

\subsubsection{Targeting Excited States with Wave-function-based Approaches}
Since electronic spectra are used to identify the oxidation states and ligand effects in actinide species, reliable theoretical predictions of excitation energies of actinide compounds are highly important. 
In CI-type methods, such as MCSCF or DMRG, the electronic excitation energies are usually obtained by calculating higher roots of the eigenvalue problem. 
However, in order to compute excited states in coupled cluster theory, we have to define a new ansatz. 
The most popular approaches applied to actinides are the equation-of-motion (EOM) and linear response (LR) coupled cluster formulations.~\cite{uranyl-6,pawel_saldien,boguslawski2016targeting,kasia-eom-pccd-erratum} 
In this chapter, we will focus on a different approach that allows us to include strong correlation effects in excited states: the Fock-space coupled cluster (FSCC) approach.  
\subsubsection{Strongly-correlated excited states with Fock-space coupled cluster theory}
The FSCC method belongs to the group of state-universal multi-reference coupled cluster theories and operates in the Fock space.  
The key idea behind the FSCC approach is to find an effective Hamiltonian in a low-dimensional model $P$ space, with eigenvalues that reliably approximate the desired eigenvalues of the real (physical) Hamiltonian. 
In the FSCC method, the $P$ space (also called the model space) contains all active valence orbitals directly involved in the electronic excitations, while the $Q$ space (also called auxiliary or complementary space) includes all remaining orbitals.
Thus, only a few eigenvalues out of the whole spectrum are calculated, reducing the expensive step of diagonalizing the Hamiltonian matrix. 
In many practical applications, however, the $P$ and $Q$ spaces are not well separated from each other, which might result in intruder state problems. 
They usually manifest as convergence difficulties for large $P$ spaces, which are particularly desired for modeling electronic structure of actinides.  
Such divergencies might occur for a specific molecule, a given molecular geometry, or basis set. 
To remedy this problem, the intermediate Hamiltonian (IH) formulation of the FSCC method has been introduced, which imposes a buffer space between the desired and undesired states.
Thus, the $P$ space is further divided into a main $P_m$ space and an intermediate $P_i$ space. 
The intermediate space serves as a buffer between the $P_m$ and $Q$ spaces, for which various numerical procedures have been developed~\cite{meissner-musial-chapter}.  

A characteristic feature of the FSCC approach is the partitioning into sectors ($k$,$l$) depending on the number of electrons removed from or attached to the reference state. Within the hole-particle formalism, the cluster operator ${T}_n$ is expressed as
\begin{equation} \label{eq:t}
{T}_n = {T}_n^{(0,0)} + {T}_n^{(0,1)} + {T}_n^{(1,0)} + {T}_n^{(1,1)} + \dots + {T}_n^{(k,l)},
\end{equation}
where the ${T}_n^{(0,0)} $ represents the ground state (zero holes and zero particles).
In the above equation, ${T}_n^{(0,1)}$ corresponds to the system with one additional electron (zero holes and one particle), ${T}_n^{(1,0)}$ reduces the number of electrons by one (one hole and zero particles), and ${T}_n^{(1,1)}$ is a single excitation (one hole and one particle).
The Hamiltonian is decomposed in the same way as the cluster operator ${T}_n$ yielding electronic energies for the individual sectors, \textit{e.g.}, the ground-state energy for sector (0,0), electron affinities for sector (0,1), ionization potentials for sector (1,0), and excitation energies for sector (1,1). 
Electronic spectra can also be obtained as a double electron attachment, that is, from sector (0,2) of the Fock space.~\cite{FSCC_02_sector,pawel3,Cristina-fscc} 
Higher order sectors have also been explored, but they are not commonly used. 
FSCC calculations require a reference determinant that dominates in the wave function expansion. 
Non-degenerate closed-shell states or high-spin open-shell states are usually the right choice for the reference determinant.

The advantage of the FSCC method is the size-extensiveness of ground-state energies and size-intensivity of excitation energies. 
The method allows us to obtain several electronic excited states of molecules with a common Fermi vacuum in a single run.
Finally, the FSCC approach includes correlation effects of core and valence electrons, while its relativistic version is appropriate for actinide-containing molecules~\cite{infante2006,uranyl-6,uran-compounds,Tecmer2014,Cristina-fscc}.

\subsubsection{Embedding Wave Function Theory in Density Functional Theory}
Reliable modeling of electronic spectra of actinide species with wave function-based methods is rather expensive and therefore usually limited to small model compounds. 
One way to overcome this problem is to combine wave function theory (WFT) with density functional theory within the so-called WFT-in-DFT approach.
Within the WFT-in-DFT framework, the whole quantum system is partitioned into a system part and into an environment part. 
The system is represented by the WFT-based method, while the environment is modeled by the (usually less accurate, but significantly less expensive) DFT approach. 
The combination of these two methods will allow us to reliably account for static and dynamic electron correlation effects in large molecular systems, yet including environmental effects at the DFT level. 
Particularly for actinides, the embedding approach allows us to account for the chemical environment, such as ligand and crystal effects, in a cost effective way.~\cite{gomes2008,pawel3,dft-func-2} 
In the simplest WFT-in-DFT embedding scheme, the DFT embedding is accounted for as a static external potential and the orthogonality between the system and environment is neglected.  
Such an embedding potential includes the electrostatic potentials of the nuclei and the electron density of the environment, as well as contributions originating from the non-additive part of the exchange--correlation energy and from the non-additive part of the kinetic energy,
\begin{align}
    v^{\text{emb}}[\rho_{\text{WFT}}, \rho_{\text{DFT}}]&= v_{\text{DFT}}^{\text{nuc}}(\mathbf{r})&  \notag  \\      
   &+\int d\mathbf{r}' \frac{\rho_{\text{DFT}}(\mathbf{r}')}{|\mathbf{r} -\mathbf{r}'|} 
   + \frac{\delta E_{\text{xc}}^{\text{nadd}} [\rho_{\text{WFT}}, \rho_{\text{DFT}}] }{\delta \rho_{\text{WFT}}}
+ \frac{\delta T_{\text{s}}^{\text{nadd}} [\rho_{\text{WFT}}, \rho_{\text{DFT}}] }{\delta \rho_{\text{WFT}}}. \\\notag
\end{align}
This one-body term is then coupled with a given WFT model.~\cite{kallay-exact-wft-in-dft} Such a WFT-in-DFT model can also be used to calculate excitation energies. 
However, the corresponding excitation spectra should be treated with care as excited states in the system might require coupling to the environment. 
\subsubsection{Interpretation of Electronic Wave Functions}\label{sec:qit}

Within molecular orbital theory, the electronic wave function is constructed from one-electron functions.
This formalism provides a convenient description of molecular systems, where the electrons occupy specific orbitals and hence are localized in certain spatial regions of molecules.
The contribution of individual orbitals to electronic structures and properties can be assessed using, for instance, concepts of quantum information theory (QIT).~\cite{dmrg-book-2,dmrg-book-3,dmrg-11,dmrg-12,entanglement_letter,dmrg-15,dmrg-16}
Specifically, QIT provides us with tools that allow us to interpret electronic wave function using the popular picture of interacting orbitals.

If a (pure) quantum state whose wave function is given by eq.~\eqref{eq:fci} cannot be written as product of states of its components (here, orbitals), $\Psi_{\rm el} \neq \psi_1 \otimes \psi_2 \otimes \dots \otimes \psi_N$, we say that the quantum state is entangled.
Thus, a single determinant wave function does not and cannot describe an entangled state.
To simplify our discussion, we will focus on a bipartite system $AB$, that is, a quantum state that is composed of two parts.
Note, however, that our analysis can be extended to quantum states that are composed of more than two subsystems $A,B,C, \dots$.
For a bipartite system $AB$, the wave function $\Psi_{\rm el}^{AB}$ of an entangled quantum state can only be written as a series of tensor products of basis states defined on the individual subsystems, $\Psi_{\rm el}^{AB} = \sum_{pq} c_{pq}\Psi_{{\rm el}, p}^{A}\otimes \Psi_{{\rm el}, q}^{B}$.
Furthermore, while the quantum state of the composite system is well-defined, the states of its components cannot be determined unambiguously, that is, the subsystems $A$ and $B$ are correlated and cannot be treated independently.
Quantum entanglement is an important feature in correlated systems such as actinide complexes and provides new perspectives on traditional quantum-chemical tools to interpret electronic structures.

A quantitative measure of the entanglement between any two subsystems is described by the von Neumann entropy and reads
\begin{equation}\label{eq:sab}
S_{A|B} = -\Tr(\rho_A \ln \rho_A),
\end{equation}
where $\rho_A$ is the reduced density matrix (RDM) for subsystem $A$ given by
\begin{equation}\label{eq:rhoa}
\rho_A = \mathrm{Tr}_B \ket{\Psi^{AB}} \bra{\Psi^{AB}}
\end{equation}
for any pure state. Thus, $\rho_A$ is obtained by tracing out all degrees of freedom from subsystem $B$ and \textit{vice versa}.
Since the von Neumann entropy corresponds to the Shannon entropy in information theory, it quantifies how much information about subsystem $A$ is encoded in subsystem $B$ and \textit{vice versa}.
The entanglement entropy eq.~\eqref{eq:sab} is determined by the eigenvalue spectrum of the RDMs.

In this chapter, we aim at quantifying the interactions between orbitals.
Thus, our subsystems should be composed of the molecular orbitals that are used to construct the Slater determinants in our wave function expansion.
For that purpose, let us rewrite the FCI wave function eq.~\eqref{eq:fci} in occupation number form (dropping the superscript)
\begin{equation}\label{eq:onv}
{\Psi}_{\rm el} = \sum_{k_1,k_2, \dots, k_L} c_{k_1,k_2, \dots, k_L} \ket{k_1,k_2, \dots, k_L},
\end{equation}
where $c_{k_1,k_2, \dots, k_L}$ are the expansion coefficients for each determinant $\ket{k_1,k_2, \dots, k_L}$ and the sum runs over all occupation number vectors in the corresponding Hilbert space.
Furthermore, we will consider only two different partitionings of our orbital space: (1) one subsystem contains exactly one orbital, while the other subsystem (here called environment) contains the remaining $L-1$ orbitals and (2) one subsystem contains exactly two orbitals, while the environment is constructed from the other $L-2$ orbitals.
More general partitioning schemes have been investigated in the literature~\cite{szalay2017correlation}, however, focusing on one- and two-orbital entanglement measures will be sufficient to provide first insights into electronic structures of molecular systems.
For the first case, we explicitly write the wave function of eq.~\eqref{eq:onv} in its bipartite form
\begin{equation}\label{eq:psiie}
\Psi_{\rm el}^{i, e}= \sum_{k_1,k_2, \dots, k_L} \tilde{c}_{k_1,k_2, \dots, k_L} \ket{k_i} \otimes \ket{e},
\end{equation}
where $\ket{e}=\ket{k_1,k_2, \dots, k_{i-1}, k_{i+1}, \dots, k_L}$ is a many-electron state vector containing environment orbitals and $\tilde{c}_{k_1,k_2, \dots, k_L}$ are the expansion coefficients that may differ from ${c}_{k_1,k_2, \dots, k_L}$ by a phase factor.
This $N$-electron state vector is then used to construct the reduced density matrix for orbital $i$, the so-called one-orbital RDM, according to eq.~\eqref{eq:rhoa} with elements
\begin{equation}\label{eq:rho1}
\rho_{i,i^\prime} = \sum_{e} \bra{e}\bra{k_i}\ket{\Psi_{\rm el}^{i, e}}\bra{\Psi_{\rm el}^{i, e}}\ket{k_{i^\prime}}\ket{e},
\end{equation}
where we sum over all many-electron states composed of the environment orbitals.
The index $i$ denotes all possible spin-occupations of a spatial orbital $i$ and includes empty orbital ($-$), doubly occupied orbitals ($\uparrow \downarrow$), orbitals with spin-up electron ($\uparrow$) and orbitals with spin-down electron ($\downarrow$).
Thus, the one-orbital RDM is a 4$\times$4 matrix and is used to calculate the entanglement entropy of a single orbital, the so-called single-orbital entropy, given by
\begin{equation}
s_i = -\sum_{\alpha=1}^4 \omega_{\alpha,i} \ln \omega_{\alpha,i},
\end{equation}
where $\omega_{\alpha,i}$ are the eigenvalues of the $i$th orbital RDM and the sum runs over all four possible occupations of a spatial orbitals.
The single-orbital entropy reaches a maximum value of ln(4).

The entanglement entropy between an orbital pair $ij$ and the remaining orbitals is obtained in a similar way.
For our second case, the environment states are defined as $\ket{e}=\ket{k_1,k_2, \dots, k_{i-1}, k_{i+1}, \dots, k_{j-1}, k_{j+1}, \dots, k_L}$, while the quantum state for this orbital partitioning reads
\begin{equation}\label{eq:psiije}
\Psi^{i, j, e}_{\rm el} = \sum_{k_1,k_2, \dots, k_L} \tilde{c}_{k_1,k_2, \dots, k_L} \ket{k_i, k_j}\otimes \ket{e}.
\end{equation}
The matrix elements of the two-orbital RDM are determined in a similar way and have the elements
\begin{equation}\label{eq:rho2}
\rho_{(i,j),(i^\prime,j^\prime)} = \sum_{e} \bra{e}\bra{k_i,k_j}\ket{\Psi^{i, j, e}_{\rm el}}\bra{\Psi^{i, j, e}_{\rm el}}\ket{k_{i^\prime},k_{j^\prime}}\ket{e}.
\end{equation}
The indices $i$ and $j$ encode all possible occupations of orbitals $i$ and $j$ in the two-orbital Fock space that is spanned by 16 states (for spatial orbitals): \mbox{($-$ $-$)}, \mbox{($\uparrow$ $-$)}, \mbox{($\downarrow$ $-$)}, \mbox{($-$ $\uparrow$)}, $\dots$, \mbox{($\uparrow \downarrow \uparrow \downarrow$)}.
Thus, the two-orbital RDM can be expressed as a 16$\times$16 matrix and determines the two-orbital entropy $s_{i,j}$~\cite{2013}.
Specifically, the two-orbital entropy quantifies the entanglement between the environment orbitals and a particular orbital pair $ij$ and is given by
\begin{equation}
s_{i,j} = -\sum_{\alpha=1}^{16} \omega_{\alpha,i,j} \ln \omega_{\alpha,i,j}	
\end{equation}
where $\omega_{\alpha,i,j}$ are the eigenvalues of the two-orbital RDM.
Given the one- and two-orbital RDMs, we can calculate the so-called mutual information between any orbital pair $ij$,
\begin{equation}
I_{i|j} = s_i + s_j - s_{i,j}.
\end{equation}
Most importantly, the mutual information is a measure of correlation and describes both classical and quantum correlations.
$I_{i|j}$ takes values in the range of $[0, \ln16]$, where 0 is obtained for uncorrelated wave functions such as a single Slater (or the HF) determinant.
We should note that evaluating the one- and two-orbital RDMs using the general eqs.~\eqref{eq:rho1} and \eqref{eq:rho2} might be cumbersome due to the phase factors that have to be accounted for in eqs.~\eqref{eq:psiie} and \eqref{eq:psiije}.
For practical calculations, the one- and two-orbital RDMs can be expressed in terms of conventional $N$-particle reduced density matrices.~\cite{dmrg-16,ijqc-eratum}
Specifically, $\rho_{i,i^\prime}$ requires only the 1- and 2-particle RDMs, while $\rho_{(i,j),(i^\prime,j^\prime)}$ requires in addition some elements of the 3- and 4-particle RDMs.
In conventional electronic structure methods, the $N$-particle RDMs are either already available or can be easily determined.
Thus, the evaluation of the single- and two-orbital entropy as well as the mutual information does not pose a computational difficulty.

The single-orbital entropy and orbital-pair mutual information are particularly useful to classify electron correlation effects into different contributions.
Large values of the entropic measures appear in molecules where strong (static and nondynamic) correlation effects dominate.
Dynamic (weak) correlation is characterized by smaller values for both $s_i$ and $I_{i|j}$, while the single-orbital entropy is close to zero for dispersion interactions.
As there is no unique definition of the different contributions to electron correlation effects, a distinction between them is rather arbitrary.
Boguslawski \textit{et al.}~\cite{dmrg-16,ijqc-eratum} proposed to dissect electron correlation effects according to the values of the single-orbital entropy and the orbital-pair mutual information which are given in Table \ref{tab:entropy}.

\begin{table}
\label{tab:entropy}
\caption{Different types of electron correlation effects and the corresponding values of the single-orbital entropy and the orbital-pair mutual information.~\cite{dmrg-16,ijqc-eratum}}
\begin{tabular}{lll}
Type of correlation         & $s_i$ & $I_{i|j}$ \\ \hline
Nondynamic                  & >0.5  & $\approx 10^{-1}$ \\
Static                      & 0.1 - 0.5  & $\approx 10^{-2}$ \\
Dynamic                     & <0.1  & $\approx 10^{-3}$ \\
Weak (dispersion, etc.)	    & $\approx 0$ & $\leq 10^{-4}$ \\ \hline
\end{tabular}
\end{table}

$s_i$ and $I_{i|j}$ can provide many additional insights in electronic structure theory calculations.
Examples are elucidating chemical bonding,~\cite{2013,dmrg-19} monitoring bond-formation processes,~\cite{Corinne_2015}, identifying transition states,~\cite{murg2015tree} and defining stable active orbital spaces in MCSCF-type calculations.~\cite{dmrg-16,ijqc-eratum,dmrg-17,boguslawski2017}
The last point may be particularly important in actinide chemistry as the number of chemically relevant orbitals is difficult to predict \textit{a priori}.
A particularly straightforward selection protocol to define stable and reliable active spaces in correlation calculations was proposed recently~\cite{entanglement_letter,dmrg-16,ijqc-eratum} and applied to plutonium~\cite{boguslawski2017} and neptunium~\cite{ola-neptunyl-cci-2018} compounds that have not been investigated using MCSCF-type approaches on a routine basis. Similar approaches have been proposed for transition metals.~\cite{stein2016automated}
The proposed selection procedure exploits the orbital-pair mutual information as sole selection criterion.
Since the orbital-pair mutual information measures orbital-pair correlations, the corresponding active orbital spaces should provide a balance description of electron correlation effects even for unknown compounds.
The protocol to obtain correlation-based active spaces includes the following steps:
\begin{enumerate}
    \item Perform a \textit{large} active space calculation with a quantum chemical method of your choice, like the DMRG algorithm, and determine the orbital-pair correlations. Note that the wave function for this large active space calculations does not have to be fully converged. For instance, in DMRG calculations, already 4-6 sweeps are sufficient to calculate the orbital-pair mutual information with sufficient accuracy
    \item Choose a cutoff threshold for the orbital-pair mutual information, for instance $I_{i|j}>10^{-2}$. Such a threshold allows us to describe static/nondynamic electron correlation
    \item Select those orbitals where $I_{i|j}$ is above the threshold
    \item Optimize the wave function for this correlation-based active space
    \item Determine the corresponding orbital-pair mutual information
    \item Compare the values of $I_{i|j}$ to those of the reference calculation in step 1. If the small active space calculation can accurately reproduce the orbital-pair correlation profile, the small active space can be considered as one \textit{optimal} choice, otherwise the acceptance threshold for $I_{i|j}$ has to be further reduced (for instance, to $10^{-3}$) and steps 3--5 have to be repeated until convergence
\end{enumerate}
Although, such correlation-based active orbital spaces represent a step towards true black-box MCSCF-type calculations, the selection criteria might have to be extended so that they allow us to consider all orbitals important for bond-breaking processes at all points of the dissociation pathway as the magnitude of orbital-pair correlations might change along the reaction coordinate.
However, technical limitations, like stability of active space calculations, cannot be excluded and will restrict all automatic active space selection protocols.

\section{Challenging Examples in Computational Actinide Chemistry}
\label{sec:3}

In the following, we briefly review some challenging case studies where computational chemistry allowed us to explain some peculiar or unexpected properties of actinide-containing compounds.

\begin{figure} \label{fig:orb}
\caption{Valence molecular orbitals and dissociation curves for the symmetric stretching of \ce{UO2^{2+}}. Subfigure (b) has been reproduced from Ref.~\cite{uranyl-dissociation} with permission from the PCCP Owner Societies.}
\centering
\includegraphics[width=0.99\textwidth]{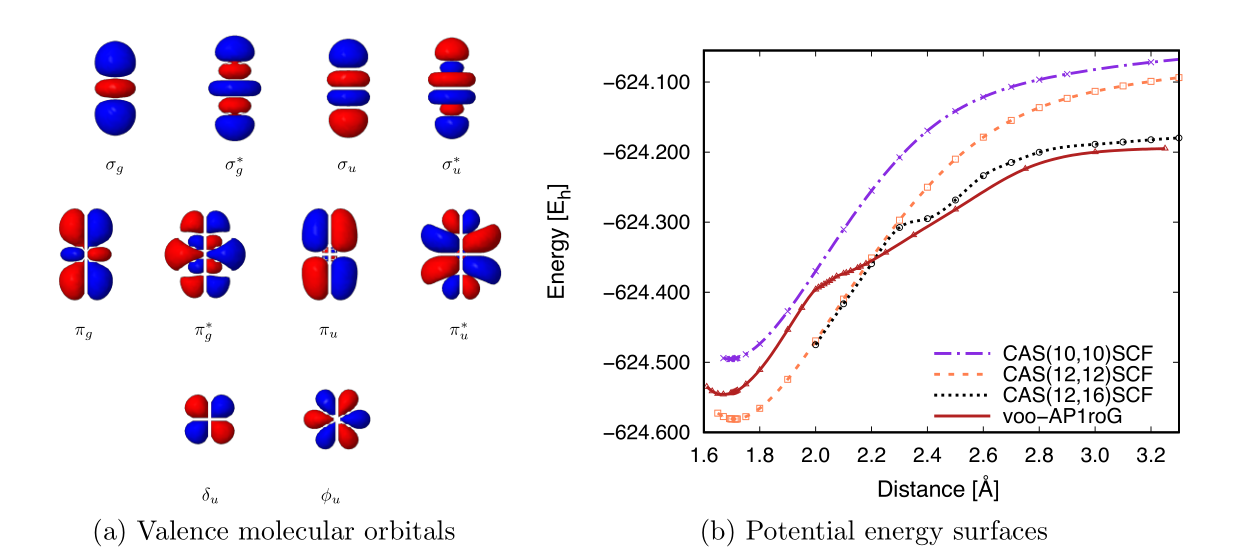}
\end{figure}

\subsection{Symmetric dissociation of UO$_2^{2+}$}
The uranyl cation is a small building block of plenty uranium-containing compounds.~\cite{denning2007,uranium_1,dft-func-2}
This molecule is characterized by a linear geometry and a singlet ground-state electronic structure.
The energetically close lying 5f, 6d, and 7s orbitals are crucial to describe the strongly-correlated valence electrons.
In addition, the uranium 6p orbitals are ``pushed from below'' by oxygen 2p electrons and thus complicate the electronic structure as 6p orbitals are easily polarizable and mix with 5f orbitals.~\cite{uranyl-dissociation,actinyl_bonding_1992,uranyl-1,neptunyl-ion-spectra-exp-2,uranyl-2,uranyl-3,dft-func-1,uranyl-5,uranyl-6,Tecmer2014}
While the bonding mechanism in UO$_2^{2+}$ is well described by single-reference CC theory for molecular structures close to the equilibrium, conventional quantum chemistry methods, like CCD, CCSD, CCSD(T), and DFT, usually fail for elongated U--O bonds.
Furthermore, the CASSCF method does not allow us to define stable and consistent active spaces along the whole dissociation pathway.
Specifically for UO$_2^{2+}$, a minimal active space (CAS(12,12)SCF) around the equilibrium should contain all $\sigma$-, $\sigma^*$-, $\pi$-, and $\pi^*$-orbitals as shown in Figure~\ref{fig:orb}(a).
For stretched U--O bonds, however, the $\phi_u$ and $\delta_u$ orbitals become partially occupied and should be included in the active space (see Figure~\ref{fig:orb}).
Since these orbitals are unoccupied around the equilibrium, they cannot be included in the active space due to convergence difficulties in CASSCF calculations.
Thus, CASSCF either predicts the wrong dissociation limit (minimal active space) or does not provide a smooth potential energy surface (CAS(12,16)SCF).
Including $\phi_u$ and $\delta_u$ orbitals into the active space results in a qualitative change in the shape of the PES featuring a shoulder around 2 \AA.
In contrast to CASSCF calculations, AP1roG allows us to include all orbitals in the active space and provides a smooth dissociation curve.
In addition, the corresponding potential energy surface features a similar shoulder as predicted by CAS(12,16)SCF.
Thus, AP1roG can capture (static/nondynamic) electron correlation effects along the dissociation pathway without imposing active spaces.


\subsection{Excitations of NUN}

The NUN complex is the isoelectronic analogue of \ce{UO2^{2+}} and has been formed in noble gas matrices and as a free molecule.~\cite{uranium_2}
This compound is particularly interesting because of its possible applications in the nuclear industry.
In the solid phase, NUN can serve as an intermediate material used in the synthesis of uranium monotride---a potential nuclear fuel.~\cite{nun_2} 
In its equilibrium geometry, the ground-state of NUN is closed-shell, similar to the isoelectronic \ce{UO2^{2+}}.
The U--N triple bonds (1.73--1.76 \AA) are slightly longer than the U--O distance in \ce{UO2^{2+}} (1.70--1.72 \AA).
In the spin-free formalism, the ground-state wave function is dominated by a single determinant (with a weight of about 0.9 for the principal determinant) with small contributions from doubly excited determinants.
The energies of the upper bonding molecular orbitals are distributed equidistantly.~\cite{uranium_2}
Furthermore, the $\delta$ and $\phi$ virtual orbitals are equally important for excited states as they lie close in energy.~\cite{uran-compounds} 
The spin--orbit electronic spectrum of NUN using different relativistic Hamiltonians is presented in Table~\ref{tbl:nun_DC}.~\cite{Tecmer2014} 
{\Large
    \begin{table}[ht!] 
     \caption{15 lowest-lying IH-FSCCSD vertical excitation energies of the NUN molecule (${\rm r}_{\rm {U-N}}=1.739$ \AA). Excitation energies are given in eV.~\cite{Tecmer2014}
}
     \label{tbl:nun_DC}
     \centering
     \resizebox{8.8cm}{!}{
\begin{tabular}{cc cc ccc}
\hline
\hline
$\Omega$ &character (from DC)&DC&DC(G)&X2C/AMF&X2C/MMF&X2C(G)/MMF\\
\hline
\hline
2$_g$&  {\scriptsize$52\%$ $\sigma_{\nicefrac{1}{2}u} \phi_{\nicefrac{5}{2}u}+26\% $ $\pi_{\nicefrac{1}{2}u} \phi_{\nicefrac{5}{2}u}$}   &0.956&0.923&0.936&0.957&0.927\\
3$_g$&  {\scriptsize$50\%$ $\sigma_{\nicefrac{1}{2}u} \phi_{\nicefrac{5}{2}u}+24\% $ $\pi_{\nicefrac{1}{2}u} \phi_{\nicefrac{5}{2}u}$}   &1.103&1.068&1.083&1.103&1.072\\
1$_g$&  {\scriptsize$45\%$ $\sigma_{\nicefrac{1}{2}u} \delta_{\nicefrac{3}{2}u}+20\% $ $\pi_{\nicefrac{1}{2}u} \delta_{\nicefrac{3}{2}u}$} &1.134&1.094&1.106&1.134&1.098\\
2$_g$&  {\scriptsize$30\%$ $\sigma_{\nicefrac{1}{2}u} \delta_{\nicefrac{3}{2}u}+15\%$ $\pi_{\nicefrac{1}{2}u} \delta_{\nicefrac{3}{2}u}$} &1.398&1.355&1.374&1.398&1.358\\
4$_g$&  {\scriptsize$49\%$ $\sigma_{\nicefrac{1}{2}u} \phi_{\nicefrac{7}{2}u}+23\%$ $\pi_{\nicefrac{1}{2}u} \phi_{\nicefrac{7}{2}u}+16\%$ $\sigma_{\nicefrac{1}{2}u} \phi'_{\nicefrac{7}{2}u}$ }   &1.699&1.645&1.680&1.698&1.646\\
3$_g$&  {\scriptsize$43\%$ $\sigma_{\nicefrac{1}{2}u} \delta_{\nicefrac{5}{2}u}+20\%$ $\pi_{\nicefrac{1}{2}u} \delta_{\nicefrac{5}{2}u}$} &1.757&1.704&1.739&1.757&1.705\\
3$_g$&  {\scriptsize$38\%$ $\sigma_{\nicefrac{1}{2}u} \phi_{\nicefrac{7}{2}u}+22\%$ $\pi_{\nicefrac{1}{2}u} \phi_{\nicefrac{7}{2}u}$ }   &2.076&2.028&2.059&2.076&2.029\\
2$_g$&  {\scriptsize$33\%$ $\sigma_{\nicefrac{1}{2}u} \delta_{\nicefrac{5}{2}u}+17\%$ $\pi_{\nicefrac{1}{2}u} \delta_{\nicefrac{5}{2}u}$} &2.519&2.476&2.502&2.519&2.478\\
1$_u$&  {\scriptsize$40\%$ $\sigma_{\nicefrac{1}{2}u} \delta_{\nicefrac{3}{2}g}+24\%$ $\pi_{\nicefrac{1}{2}u} \delta_{\nicefrac{3}{2}g}$ } &2.669&2.696&2.680&2.669&2.696\\
0$^+_u$&  {\scriptsize$41\%$ $\sigma_{\nicefrac{1}{2}u} \sigma_{\nicefrac{1}{2}g}+30\%$ $\pi_{\nicefrac{1}{2}u} \sigma_{\nicefrac{1}{2}g}$ } &2.709&2.757&2.740&2.709&2.755\\
1$_u$&  {\scriptsize$41\%$ $\pi_{\nicefrac{3}{2}u} \sigma'_{\nicefrac{1}{2}g}+29\%$ $\pi'_{\nicefrac{3}{2}u} \sigma''_{\nicefrac{1}{2}g}$ }    &2.711&2.759&2.743&2.711&2.757\\
1$_g$&  {\scriptsize$80\%$ $\pi_{\nicefrac{3}{2}u} \phi_{\nicefrac{5}{2}u}$}      &2.711&2.675&2.690&2.711&2.679\\
2$_u$&  {\scriptsize$34\%$ $\sigma_{\nicefrac{1}{2}u} \delta_{\nicefrac{3}{2}g}+24\%$ $\pi_{\nicefrac{1}{2}u} \delta_{\nicefrac{3}{2}g}$ }    &2.749&2.767&2.758&2.749&2.768\\
4$_g$&  {\scriptsize$84\%$ $\pi_{\nicefrac{3}{2}u} \phi_{\nicefrac{5}{2}u}$}      &2.844&2.808&2.823&2.844&2.811\\
2$_u$&  {\scriptsize$73\%$ $\sigma_{\nicefrac{1}{2}g} \phi_{\nicefrac{5}{2}u}$}   &2.895&2.857&2.875&2.895&2.860\\
\hline
\hline
\end{tabular}
}
\end{table}
}
DC denotes the standard Dirac--Coulomb Hamiltonian, DC(G) is the DC Hamiltonian augmented with the Gaunt operator at the SCF level, X2C/AMF and MMF correspond to the X2C Hamiltonian with the atomic and molecular mean field approximations to spin--orbit coupling, and X2C(G) is again the X2C Hamiltonian augmented with the Gaunt operator at the SCF level.
Based on the data presented in Table~\ref{tbl:nun_DC}, we can conclude that NUN possesses significant multi-reference character and a rather complex electronic spectrum.~\cite{Tecmer2014}
Furthermore, including the Gaunt operator in the Hamiltonian has only negligible effect on the electronic spectra of NUN, while the X2C Hamiltonian represents a computationally cheaper alternative to the full DC Hamiltonian. 
Specifically, the spin--orbit electronic spectrum calculated within X2C/MMF is almost identical to the DC spectrum. 


\subsection{CUO Diluted in Noble Gas Matrices}

As noble gases are known to be inert, they constitute an ideal environment to investigate properties of single molecules.
Diluted in a noble gas matrix, the studied substance should not interact with the environment and thus its electronic structure should remain unaffected.
A peculiar observation has been made for the CUO molecule and was reported by Andrews \textit{et al}.~\cite{cuo_1,cuo_2,cuo_3}
Experiments revealed a blue shift in the asymmetric U--O and U--C vibrational spectra when the composition of the noble gas matrix---a mixture of neon and argon---was changed.
To explain the observed shifts, experimentalist anticipated that the noble gas environment may not be inert and may interact with the CUO unit differently depending on its composition.


\begin{figure}[b] \label{fig:spinsplittings}
\caption{Spin state energy splittings of CUO diluted in different noble gas matrices. All calculations have been performed using the spin-free (SF) DKH Hamiltonian of 10th order, while a spin-orbit (SO) correction was added \textit{a posteriori}. $^3\Phi_{(v)}$: vertically excited $^3\Phi$ state. $^3\Phi_{(a)}$: adiabatically excited $^3\Phi$ state. Reproduced from Ref.~\cite{cuo_dmrg} with permission from the PCCP Owner Societies.}
\centering
\includegraphics[width=0.99\textwidth]{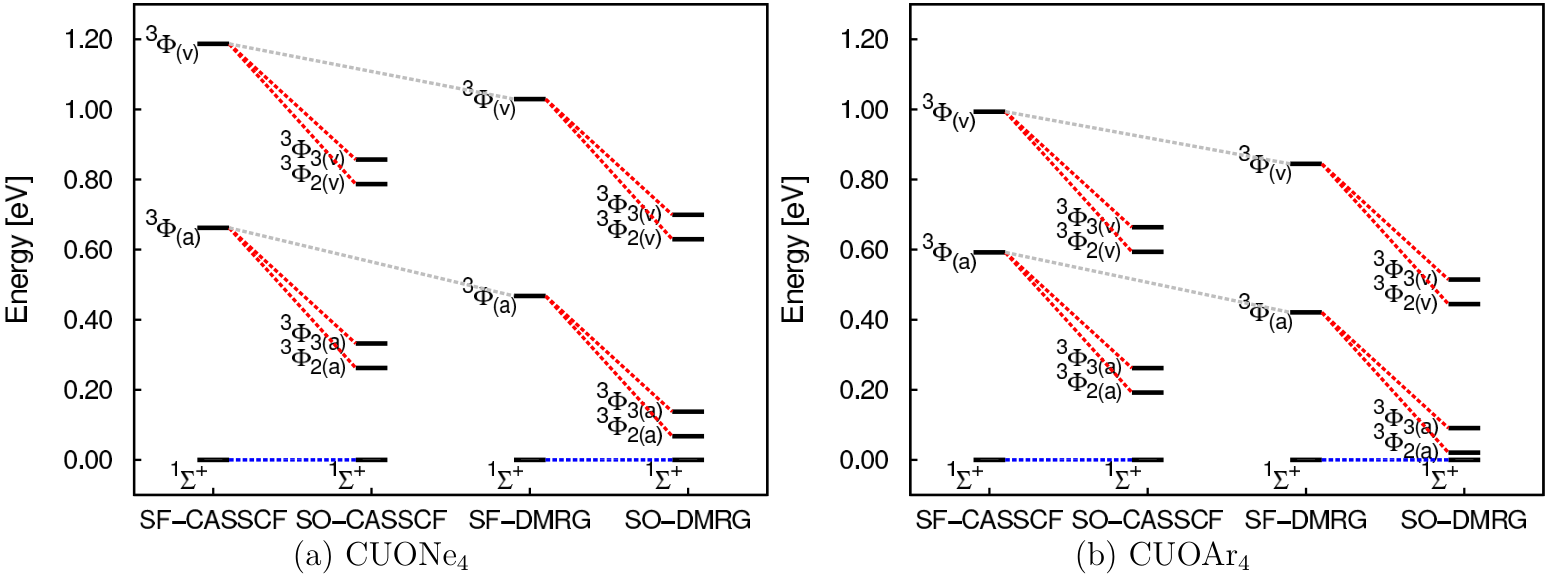}
\end{figure}

To understand the specific interaction between CUO and noble gas matrices, various quantum chemistry methods have been applied both to the bare CUO unit and to small model systems including noble gas atoms.
A new perspective has been provided by \textit{ab initio} calculations using scalar-relativistic high-level wave-function based methods with spin-orbit corrections added \textit{a posteriori}.~\cite{cuo_dmrg}
A computational study covering spin-orbit (SO)-CASSCF and SO-DMRG calculations indicate that the interaction of CUO with Ne atoms does not change the order of states and the ground state remains a $^1\Sigma^+$ state.
However, the energy gap between the two lowest-lying states becomes negligible in \ce{CUOAr4} and the molecule requires a multi-reference treatment.
Ref.~\cite{cuo_dmrg} was the first work dissecting electron correlation effects using quantum information theory measures in \ce{CUONe4} and \ce{CUOAr4} molecules. The interactions between CUO and Ar valence electrons has been confirmed supporting experimental and theoretical anticipations. Furthermore, numerical results suggest that a (thermal) spin crossover may occur.

\subsection{Cation--cation Attraction in [NpO$_2$]$_2^{2+}$}

\begin{figure}[b]
\caption{(a) Molecular geometries of neptunyl CCIs including explicit water molecules and (b) orbital-pair correlations of the diamond-shaped (bottom) [NpO$_2$]$_2^{2+}$. The values of the single-orbital entropy are coded by the size of the dots corresponding to each orbital. The strongest correlated orbitals are connected by blue lines ($I_{i|j}>10^{-1}$), followed by orbital-pair correlations marked by red lines ($10^{-1}>I_{i|j}>10^{-2}$).}\label{fig:cci}
\centering
\includegraphics[width=1.0\textwidth]{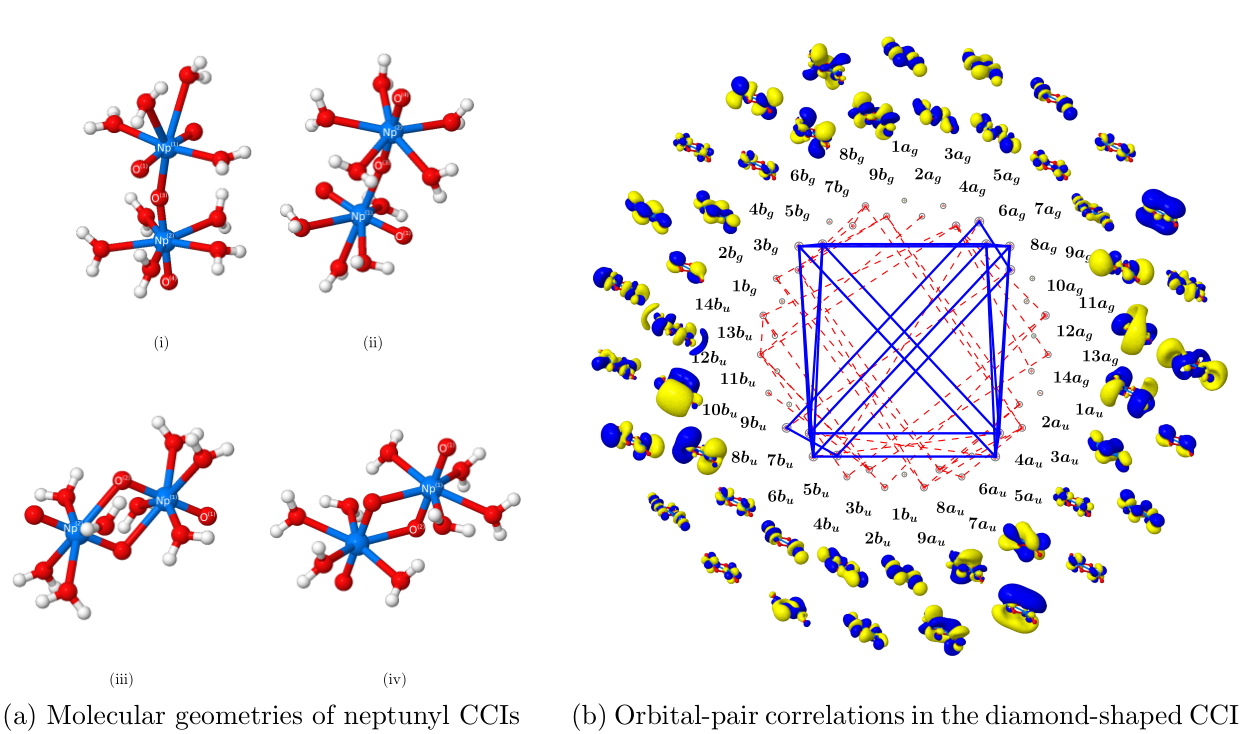}
\end{figure}

The attractive interaction between two cations is a characteristic feature of heavy-element-containing molecules and has been observed for the first time in uranyl perchlorate solution and aqueous chlorate media.~\cite{sullivan1960, sullivan1961, sullivan1962}
This so-called cation--cation interactions (CCI) are exploited in the synthesis of new crystalline structures.~\cite{Arnold2009, Jin2011, Jin2011-2, wang2011, wang2012, diwu2012, laura2013, ccis-in-solids-1, ccis-in-solids-2, ccis-in-solids-3}
Most importantly, CCIs pose a technical difficulty when reprocessing spent nuclear fuel. 

Another example for this peculiar interaction are neptunyls.
Specifically, neptunyl CCIs feature an end-on or side-on interaction producing the T-shaped or diamond-shaped dimers.
The CCIs structures are stable primarily because of the bonding interaction between the oxygen and the neptunium atoms of two neighbouring complexes, where the effective charge of the oxygen atoms is negative \cite{vallet2004} in contrast to the effective positive charge localised on the actinide atoms.~\cite{choppin1984}
The stability of CCIs is strongly influenced by the \mbox{Np--O} bond distance, \cite{guillaume1981} which changes in different environments.~\cite{rao1979, neptunium_spectra_1, halperin1983, roesch1990}
In order to describe the ground- and excited-state properties of such CCIs, the theoretical model needs to account for environmental effects originating from the solvent.
Figure~\ref{fig:cci}(a) shows two different explicit solvation models for the T-shaped (i and ii)  and diamond-shaped (iii and iv) clusters.
Including both explicit and implicit solvation models in quantum chemistry calculations allows us to reproduce the experimentally measured Np--Np distance within (spin-free) DFT calculations.
However, in order to accurately model the ground state (and also excited state) electronic structure, we have to account for both relativistic and correlation effects on an equal footing.
This poses a particular challenge for conventional electronic structure methods primarily because we have to deal with two heavy-element centers.

The strong multi-reference nature of the neptunyl CCIs becomes evident when we perform an orbital correlation analysis.
The orbital-pair mutual information for the ground-state of the diamond-shaped neptunyl CCI \ce{[NpO2]2^{2+}} is shown in Figure~\ref{fig:cci}(b).
The correlation between orbital pairs is indicated by lines, while its strength is color-coded: strong correlations are shown in blue, medium-sized correlations in red, etc.
Specifically for the diamond-shaped \ce{[NpO2]2^{2+}}, the $\sigma_g$- and $\sigma_g^*$-type orbitals are as important as $\delta_u$- and $\phi_u$-type orbitals.
Note that $\pi$-type orbitals are only moderately correlated with each other.
The orbital-correlation analysis, thus, suggests that a balanced active space for neptunyl-containing CCIs that allows us to describe both nondynamic and static correlation (or moderately and strongly correlated orbitals) should contain approximately 30 orbitals ($\delta_u$-, $\phi_u$-, bonding and antibonding combinations of $\sigma_g$-, $\sigma_u$-, $\pi_u$-, and $\pi_g$-type orbitals of each monomer).
However, such large active spaces are difficult to handle with conventional multi-reference methods.
In a first approximation, we can consider active spaces where only the strongest orbital-pair correlations are accounted for, while the remaining correlations are treated \textit{a posteriori} using, for instance, perturbation theory.
Such a study has been presented recently in Ref.~\cite{ola-neptunyl-cci-2018} and highlights the interplay of relativistic and correlations effects in neptunyl CCIs.
 

\section{Summary}
\label{sec:4}

This chapter reviewed different quantum chemistry approaches applicable to actinide chemistry.
Actinide-containing compounds are particularly challenging as both relativistic effects and correlation effects have to be accounted for on an equal footing.
Specifically, we have focused on the most common relativistic Hamiltonians and wave-function-based methods that allow us to reliably model actinide chemistry.
Particularly interesting are novel and unconventional methods, like the DMRG algorithm or geminal-based approaches as they allow us to include a large number of orbitals in active space calculations.
This feature is desirable especially for multi-centered actinide-containing compounds.

Our numerical examples show the strengths and pitfalls of present-day quantum chemistry methods when the systems under study contain one or more heavy-elements.
While DFT can accurately provide molecular geometries, wave-function-based methods have to be applied to describe electronic structures of ground- and excited-states.
Furthermore, conventional methods like CASSCF fail in describing potential energy surfaces for stretched actinide--ligand bonds.
Such calculations can only be accomplished using modern and unconventional methods like DMRG or AP1roG.

\begin{acknowledgement}
A.~\L.~and K.~B.~acknowledge financial support from the National Science Centre, Poland (SONATA BIS 5 Grant No.~2015/18/E/ST4/00584).
K.~B.~gratefully acknowledges funding from a Marie-Sk\l{}odowska-Curie Individual Fellowship project no.~702635--PCCDX and a scholarship for outstanding young scientists from the Ministry of Science and Higher Education.
P.~T.~thanks the POLONEZ fellowship program of the National Science Center (Poland), No. 2015/19/P/ST4/02480. This project had received funding from the European Union's Horizon 2020 research and innovation programme under the Marie Sk{\l}odowska--Curie grant agreement No. 665778.
\end{acknowledgement}

\bibliographystyle{spbasic}
\bibliography{rsc}                                                                                            
\end{document}